\documentclass[english,prd,nofootinbib,preprintnumbers,twocolumn,showpacs]{revtex4}
\usepackage[latin1]{inputenc}
\usepackage{amsmath}
\usepackage{amsfonts}
\usepackage{appendix}
\usepackage{amssymb}
\usepackage{epsfig}
\usepackage{graphics,psfrag,rotating}
\usepackage{graphicx}
\usepackage{dcolumn}
\usepackage{bm}
\usepackage{epstopdf}
\usepackage{color}
\usepackage[usenames,dvipsnames,svgnames]{xcolor}
\usepackage[colorlinks=true,
            linkcolor=red,
            urlcolor=gray,
            citecolor=blue]{hyperref}
\usepackage{hyperref}
\usepackage[T1]{fontenc}
\usepackage{multirow}
\usepackage{float}
\usepackage{cancel}
\usepackage{hyperref}
\usepackage{adjustbox}
\usepackage{subfigure}

\usepackage{enumitem}

\def\3nab{\tilde{\nabla}}

\def\be {\begin{equation}}
\def\ee {\end{equation}}
\def\ba {\begin{align}}
\def\ea {\end{align}}

\def\bc {\begin{center}}
\def\ec {\end{center}}
\def\case#1/#2{\frac{#1}{#2}}

\newcommand{\bver}{\begin{verbatim}}
\newcommand{\ever}{\end{verbatim}}

\newcommand{\bea}{\begin{eqnarray}}
\newcommand{\eea}{\end{eqnarray}}
\newcommand{\beaa}{\begin{eqnarray*}}
\newcommand{\eeaa}{\end{eqnarray*}}

\def\case#1/#2{\textstyle\frac{#1}{#2}}

\begin{document}

\title{Modified gravity or imperfect dark matter: a model-independent discrimination}

\author{
Miguel Aparicio Resco 
}
\email{migueapa@ucm.es}
\affiliation{Departamento de F\'{\i}sica Te\'orica and Instituto de F\'isica de Part\'iculas y del Cosmos (IPARCOS), Universidad Complutense de Madrid, 28040 Madrid, Spain}
\author{Antonio L.\ Maroto}
\email{maroto@ucm.es}
\affiliation{Departamento de F\'{\i}sica Te\'orica and Instituto de F\'isica de Part\'iculas y del Cosmos (IPARCOS), Universidad Complutense de Madrid, 28040 Madrid, Spain}

\pacs{04.50.Kd, 98.80.-k, 98.80.Cq, 12.60.-i}



\begin{abstract}
We analyze how to parametrize  general modifications of the dark matter perturbations equations in a model-independent way. We prove that a general model with an imperfect and non-conserved  dark matter fluid with bulk and shear viscosities and heat flux in a modified gravity scenario can be described with five general functions of time and scale. We focus on the sub-Hubble regime within the quasi-static approximation and calculate the observable power spectra of the galaxy distribution, galaxy velocities and weak lensing and find that these observables are only sensitive to three combinations of the initial five functions. Deviations of these three observable functions with respect to $\mathrm{\Lambda CDM}$ give us different characteristic signals which allow us to determine in which cases it is possible to discriminate a modification of gravity from an imperfect or non-conserved dark matter. Finally, we perform a Fisher forecast analysis for these three parameters and show an example for a particular model with shear viscosity.

\end{abstract} 

\maketitle

\section{Introduction} \label{sec0}

The $\mathrm{\Lambda CDM}$ model has become the current concordance model thanks to its simplicity and its excellent agreement with most of the  observational data, from the Cosmic Microwave Background (CMB) \cite{Adam:2015rua, Ade:2015rim} to the accelerated expansion of Universe via the Supernovae measurements \cite{Perlmutter:1998np, Riess:1998cb} or the large-scale structure (LSS) observables from galaxy maps \cite{Dawson:2012va, Anderson:2013zyy, Alam:2016hwk}. However, to date $\mathrm{\Lambda CDM}$ can only be considered as a phenomenological description rather than a fundamental theory since the nature of its dark sector remains completely unknown. This fact has motivated the search for underlying theories which could shed new light on the properties of such sector. Among the various possibilities that have been explored, more general gravity theories based on well-motivated fundamental physical  principles \cite{Weinberg:1988cp, Peebles:2002gy} and models with additional degrees of freedom have received much attention in recent years. In this line, theories like quintessence \cite{Copeland:2006wr, Zlatev:1998tr},  $f(R)$ \cite{DeFelice:2010aj, Starobinsky:2007hu, Hu:2007nk} and many others \cite{DE} can fit observations, including the accelerated expansion of Universe with the same precision as a cosmological constant. 

On the other hand, although $\Lambda$CDM assumes a perfect non-relativistic fluid for dark matter  which is decoupled from baryons and dark energy i.e. cold dark matter (CDM), many other possibilities cannot be a priori excluded. Indeed,  given our limited observational information on the dark components, additional interactions within the dark sector that could modify the background and the evolution of CDM perturbations have been proposed \cite{Zimdahl:2001ar, Farrar:2003uw}. As a matter of fact, it has been shown  that these interacting model could alleviate some of the problems of the $\mathrm{\Lambda CDM}$ model such as the so called coincidence problem \cite{Chimento:2003iea}. In addition, there are some experimental tensions between the CMB and local observables which could be reduced in modified scenarios  \cite{Riess:2016jrr,Valentino}. 
On the other hand, there are also some well-known problems related to standard CDM at sub-galactic scales such as the problem of missing satellites \cite{Moore:1999nt, Klypin:1999uc} or the cusp-core problem \cite{deBlok:2009sp}, but also on larger scales, as suggested by  the Planck measurements that observed less clusters than expected \cite{Ade:2015fva}, which could point towards a  modification of  the perfect-fluid CDM scenario. In this sense, in addition to the high variety of dark energy and modified gravity models, there are also several proposals for imperfect and non-conserved dark matter. These models can be 
classified into three main types: interacting  dark sector models \cite{Wang:2016lxa, Tamanini:2015iia}, models of an imperfect fluid with bulk and shear viscosity \cite{Barbosa:2017ojt, Velten:2011bg}, and models of an imperfect fluid with heat flux. There are different  approaches for interacting models but in most of them, the interaction term is assumed proportional to the dark matter or dark energy densities \cite{Yin:2015pqa}. These interacting models have been proved to be compatible with current observations \cite{He:2008si, He:2009mz}. On the other hand, imperfect fluid models for dark matter have been considered as possible solutions to the small-scale problems of CDM  \cite{Barbosa:2017ojt}. Regarding  models with bulk viscosity, it has been shown that they can generate a negative pressure contribution that can accelerate the universe expansion  \cite{Kremer:2002hz, Wilson:2006gf}. Nevertheless, if we consider it as the only contribution to the late-time acceleration, we find problems and it seems necessary to include dark energy \cite{Piattella:2011bs, Giovannini:2015uia}. Finally, practically there are no models of dark matter with heat flux proposed in the literature as in the standard picture dark matter has to be mainly cold. However a general dark matter model could have some heat flux effect and for completeness we will consider it in the general scenario.

On the experimental counterpart, future generations of galaxy surveys like J-PAS \cite{Benitez:2014ibt}, DESI \cite{Aghamousa:2016zmz}, Euclid \cite{Laureijs:2011gra} or SKA  \cite{Johnston:2008hp} will improve the accuracy of cosmological measurements and could reaffirm or disprove the $\mathrm{\Lambda CDM}$ model. In these surveys, the main observables are the power spectra of galaxy density, peculiar velocities and weak lensing convergence. The galaxy density and peculiar velocities power spectra are sensitive to the growth of CDM perturbations, and also to the background cosmology via the Alcock-Paczynski effect \cite{Alcock:1979mp}. The weak-lensing effect \cite{Bacon:2000sy, Kaiser:2000if} is the distortion of the galaxy shapes due to the gravitational perturbations.  In addition to the information on the growth of dark matter perturbations, the weak lensing measurements provide information on the gravitational slip which is related to the gravitational potential that light sees \cite{DE}.

To model these power spectra for the galaxy and lensing maps, two useful approximations are usually taken into account: the sub-Hubble approximation, and the quasi-static approximation (QSA). With this last approximation, we neglect time derivatives of perturbations with respect to  the spatial derivatives in the sub-Hubble regime. Assuming these approximations, it is well-known that a general modified gravity model with extra scalar degrees of freedom can be described with only two additional parameters: an effective Newton constant $\mu = G_{eff}/G$ and a gravitational slip parameter $\gamma$ \cite{Pogosian:2010tj, Silvestri:2013ne}. However, when including the possibility of having imperfect and non-conserved dark matter, this simple effective description in no longer valid. In this work we analyze how to  characterize a non-standard dark matter fluid  in a model-independent way. We find that a total of five parameters are needed to characterize a general modification of gravity and dark matter. We obtain the effects of this parametrization in the observable power spectra and find that, although we need five independent parameters to describe the full theory, the observables are only sensitive to three different combinations of them.
We conclude that, even at the perturbation level, the observable effects of a modification of gravity can be mimicked by a modification of the dark matter properties in certain cases. 
However, we will also show that, in other cases, it is possible to distinguish them  observationally.  Finally, we perform a  Fisher matrix forecast for these three parameters using Euclid and SKA-WALLABY surveys and analyze a particular model with shear viscosity.

The paper is organized as follows: in \ref{sec1} we review the standard case for modified gravity with the $\mu$ and $\gamma$ parameterization for scalar degrees of freedom. Then in \ref{sec2} we generalize the conservation equations, we define the parameters and finally we obtain their general expressions. In section \ref{sec3}, we consider a non-conserved and imperfect fluid for dark matter. We obtain expressions for the parameters in different cases: fluids with bulk and shear viscosity, fluids with heat flux and a non-conserved fluid. In section \ref{sec4} we obtain analytic approximations for the growth function in terms of the new parameters. We use these  results in \ref{sec5} to obtain the effect of this parameterization in the observable power spectra of dark matter density contrast, peculiar velocities and convergence power spectrum. In \ref{sec6} we perform a Fisher analysis for the three observable parameters in Euclid and SKA-WALLABY surveys; and in \ref{sec7} we constraint a particular model with shear viscosity using SDSS luminous red galaxies. Finally in \ref{sec8} we discuss the results and conclusions.   

\section{Modified gravity phenomenological parametrization} \label{sec1}

In this section we review the phenomenological parametrization of modified gravity presented in \cite{Silvestri:2013ne}. We consider a modification of General Relativity in the  presence of extra scalar degrees of freedom $\phi_i$. We start with the scalar perturbation of the flat Robertson-Walker metric in the longitudinal gauge,

\begin{align}\label{s1.1}
ds^2 = a^2(\tau) [-(1+2\Psi)d\tau^2 + (1-2\Phi)d{\bf x}^2],
\end{align}
where $\tau$ is the conformal time. Adittionaly, we consider the perturbation of each scalar degree of freedom $\phi_i = \phi^0_i (\tau) + \delta \phi_i (\tau, \bf x)$. Then the modified Einstein equations at the perturbation level are,

\begin{equation}\label{s1.2}
\delta \bar{G}^{\mu}_{\,\,\,\nu}=8 \pi G \, \delta T^{\mu}_{\,\,\,\nu},
\end{equation}
where the perturbed modified Einstein tensor $\delta \bar{G}^{\mu}_{\,\,\,\nu}$  can depend on both the metric potentials
$\Phi$, $\Psi$ and the perturbed fields $\delta\phi_i$ to first order. The only matter-energy content relevant at late times is pressureless matter so that,
\begin{equation}\label{s1.3}
\delta T^{0}_{\,\,\,0}=-\rho \, \delta,
\end{equation}
\begin{equation}\label{s1.4}
\delta T^{0}_{\,\,\,i}=-\rho \, v_{i},
\end{equation}
\begin{equation}\label{s1.5}
\delta T^{i}_{\,\,\,j}=0,
\end{equation}
where $v_{i}$ is the three-velocity of matter, $\rho$ is the density and $\delta$ the density contrast. A priori, we can construct four independent equations but, as shown in \cite{Resco:2018ubr}, due to the Bianchi identities we have only two: $\delta \bar{G}^{0}_{\,\,0}$ and $\delta \bar{G}^{i}_{\,\,i}$. At first order in perturbations they read,

\begin{eqnarray}\label{s1.6}
a_{1 1} \, \Psi + a_{1 2} \, \Phi + \sum_{i=1}^N a_{1 i+2} \, \delta \phi_i = -8 \pi G a^2\, \rho \, \delta,
\end{eqnarray}
\begin{eqnarray}\label{s1.7}
a_{2 1} \, \Psi + a_{2 2} \, \Phi + \sum_{i=1}^N a_{2 i+2} \, \delta \phi_i = 0,
\end{eqnarray}
where $N$ is the number of scalar degrees of freedom, $a_{i j}$ are general differential operators that we will restrict to be of second order. Now we introduce the quasi-static approximation in which we neglect all time derivatives of perturbations. Then, if we consider the equations in Fourier space, equations (\ref{s1.6}) and (\ref{s1.7}) become algebraic equations and $a_{i j}$  functions of time and scale $k$. For each scalar degree of fredoom we would have an equation and using them we could obtain in the quasi-static approximation the following expressions,

\begin{eqnarray}\label{s1.8}
\delta \phi_i = b_{1i} \Psi + b_{2i} \Phi,
\end{eqnarray}
being $b_{i j}$ functions of time and scale $k$. Using equations (\ref{s1.8}) we can solve the system (\ref{s1.6}-\ref{s1.7}) and  write the solution as,

\begin{eqnarray}\label{s1.12}
k^2 \, \Phi = -4\pi G\,a^2 \, \mu \, \gamma \, \rho\, \delta,
\end{eqnarray}
\begin{eqnarray}\label{s1.13}
k^2 \, \Psi = -4\pi G\,a^2 \, \mu\, \rho \,\delta,
\end{eqnarray}

where,

\begin{eqnarray}\label{s1.14}
\gamma = - \frac{a_{21} + \sum_{i=1}^N a_{2 i+2} b_{1i}}{a_{22} + \sum_{i=1}^N a_{2 i+2} b_{2i}},
\end{eqnarray}
\begin{eqnarray}\label{s1.15}
\mu = \frac{2 k^2}{a_{11} + \sum_{i=1}^N a_{1 i+2} b_{1i} + \gamma \left( a_{12} + \sum_{i=1}^N a_{1 i+2} b_{2i} \right)}, \nonumber\\
\end{eqnarray}
If we assume the  coefficients $a_{ij}$ and $b_{ij}$ to be quadratic in $k$, we recover the explicit expressions for $\mu$ and $\gamma$ obtained in  \cite{Silvestri:2013ne}. Here, we remark the fact that, in the quasi-static approximation, a general modification of gravity with additional scalar degrees of freedom can be characterized by two functions $\mu(k,a)$ and $\gamma(k,a)$ (see \cite{Resco:2018ubr, vector2} for the vector case).

Finally, if we want to complete the problem we need the dark matter conservation equations. In the standard case the conservation equations read,

\begin{eqnarray}\label{s1.16}
\nabla_{\mu} {T}^{\mu}_{\,\,\nu}=0,
\end{eqnarray}

For a pressureless matter we have,

\begin{eqnarray}\label{s1.17}
 T^{\mu}_{\,\,\,\nu}=\rho\, u^{\mu}u_{\nu},
\end{eqnarray}

being

\begin{eqnarray}\label{s1.18}
\rho=\rho_0+\delta\rho,
\end{eqnarray}

and the four-velocity of matter $u^\mu=dx^{\mu}/ds$ is

\begin{eqnarray}\label{s1.19}
u^\mu=a^{-1}(1-\Psi,v^i),
\end{eqnarray}

so that

\begin{eqnarray}\label{s1.20}
u_\mu=a(-1-\Psi,v_i).
\end{eqnarray}

Because we are considering only scalar perturbations, the velocity perturbation is longitudinal so that $\hat{v}=\hat{k}$. Then we can obtain from (\ref{s1.16}) two scalar equations: $\nabla_{\mu} {T}^{\mu}_{\,\,0}=0$ and $k^i \, \nabla_{\mu} {T}^{\mu}_{\,\,i}=0$. If we apply the quasi-static approximation we readily obtain,

\begin{eqnarray}\label{s1.21}
\delta' = -\theta,
\end{eqnarray}
and
\begin{eqnarray}\label{s1.22}
\theta' = - {\cal H} \theta + k^2 \Psi,
\end{eqnarray}
where prime denotes derivative with respect to $\tau$, ${\cal H} = a' / a$ is the comoving Hubble parameter and $\theta = i k_i v^i$. If we derive equation (\ref{s1.21}) with respect to $\tau$ and use equations (\ref{s1.21}) and (\ref{s1.22}) we can obtain the evolution  equation for the density contrast $\delta$,
\begin{eqnarray}\label{s1.23}
\delta'' + {\cal H} \, \delta' + k^2 \Psi = 0,
\end{eqnarray}
Using equation (\ref{s1.13}) in (\ref{s1.23}) we obtain,

\begin{eqnarray}\label{s1.24}
\delta'' + {\cal H} \, \delta' - \frac{3}{2} \, {\cal H}^2 \, \Omega_m (a) \, \mu \, \delta = 0,
\end{eqnarray}
where we have used $4\pi G\,a^2 \, \rho = \frac{3}{2} \, {\cal H}^2 \, \Omega_m (a)$, being $\Omega_m (a) = \rho_m (a) / \rho_c (a)$, $\rho_c (a) = 3 H^2(a) / 8\pi G$ and $H(a) = a^{-1} {\cal H}(a)$ the Hubble parameter. 

To summarize we have considered modified gravity  equations (\ref{s1.6}-\ref{s1.7}) tha can be characterized by $\mu$ and $\gamma$ functions, together with the standard conservation  equations for dark matter (\ref{s1.21}-\ref{s1.22}). In next section we will extend this formalism for the case in which we also modify the dark matter conservation equations in the most general way.

\section{Phenomenological parametrization of dark matter growth} \label{sec2}

Now we want to modify the conservation equations (\ref{s1.21}-\ref{s1.22}) in a general way. We will consider also a general modification of gravity equations (\ref{s1.6}-\ref{s1.7}) which can be encoded in $\mu$ and $\gamma$ parameters via equations (\ref{s1.12}-\ref{s1.13}). Because the scalar perturbations are $(\theta,\delta,\Psi,\Phi)$, the most general way in which we can modify (\ref{s1.21}-\ref{s1.22}) is,

\begin{eqnarray}\label{s2.1}
\delta' = - c_{11} \, \theta + c_{12} \, {\cal H} \delta + c_{13} \, {\cal H} \Psi + c_{14} \, {\cal H} \Phi,
\end{eqnarray}
\begin{eqnarray}\label{s2.2}
\theta' = - c_{21} \, {\cal H} \theta + c_{22} \, {\cal H}^2 \delta + c_{23} \, k^2 \Psi + c_{24} \, k^2 \Phi,
\end{eqnarray}

With this parametrization $c_{ij}$ are in general dimensionless functions of time and scale, and we recover the standard case when $c_{11} = c_{21} = c_{23} = 1$ and $c_{ij} = 0$ for the rest of $i, j$. We could have other scalar degrees of freedom with perturbations $\delta q_i$ but, in this situation, we would also have equations for those degrees of fredoom and we could find the relations $\delta q_i = \delta q_i(\theta,\delta,\Psi,\Phi)$. Then we can always find equations of the form of (\ref{s2.1}-\ref{s2.2}).

As in the previous section, we derive equation (\ref{s2.1}) with respect to conformal time.  Using (\ref{s2.1}) and (\ref{s2.2}), and considering the modified gravity equations (\ref{s1.12}-\ref{s1.13}) we obtain,

\begin{eqnarray}\label{s2.3}
\delta'' + {\cal H} \, \mu_d \, \delta' - \frac{3}{2} \, {\cal H}^2 \, \Omega_m (a) \, \mu_m \, \delta = 0,
\end{eqnarray}

being,

\begin{align}\label{s2.4}
\mu_m = c_{11} \, \mu \, & \left( c_{23} - \frac{{\cal H}^2}{k^2} \, {\cal C}_{3} + \gamma \, \left[ c_{24} - \frac{{\cal H}^2}{k^2} \, {\cal C}_{4} \right] \right) \nonumber\\ 
&- \frac{2 c_{11}}{3 \Omega_m (a)} \, \left( c_{22} - {\cal C}_{2} \right),
\end{align}
\begin{eqnarray}\label{s2.5}
\mu_d = c_{21} - c_{12} - \frac{c_{11}'}{{\cal H} \, c_{11}},
\end{eqnarray}

with,

\begin{eqnarray}\label{s2.6}
{\cal C}_{i} \equiv \frac{c_{1i}}{c_{11}} \, \left[ c_{21} - \frac{c_{11}'}{{\cal H} \, c_{11}} + \frac{{\cal H}'}{{\cal H}^2} + \frac{c_{1i}'}{{\cal H} \, c_{1i}} \right].
\end{eqnarray}

As we can see, if we want to parameterize the density contrast evolution, we need only two independent parameters $(\mu_m, \mu_d)$. On the other hand,  to obtain the velocity perturbation $\theta$ as a function of the matter density contrast $\delta$ we can always rewrite equation (\ref{s2.1}) in the following form,

\begin{eqnarray}\label{s2.7}
\theta = - \mu_{\theta} \, \delta',
\end{eqnarray}
where,
\begin{align}\label{s2.8}
\mu_{\theta} = \frac{1}{c_{11}} - \frac{{\cal H} \, \delta}{c_{11} \, \delta'} \, \left[ c_{12} - \frac{3}{2} \, \frac{{\cal H}^2}{k^2} \, \Omega_m (a) \, \mu \, (c_{13} + \gamma \, c_{14}) \right]. \nonumber\\
\end{align}
Here $\delta$ and $\delta'$ are obtained from the solutions of
\eqref{s2.3}. 
Thus we see that in order to describe the general modified system of equations for matter and gravity perturbations, we need in total five effective parameters $(\mu, \gamma, \mu_m, \mu_d, \mu_{\theta})$. Notice that now, the presence of an imperfect dark matter implies that, in general, the effective Newton constant that controls the growth of matter perturbations given by $\mu_m$ may be different from  the effective constant that light sees, which is given by the combination $\gamma(1+\mu)/2$, even when $\gamma = 1$.

In the following section we will prove that a very broad class of models can be parameterized with these five parameters.

\section{Non-conserved and imperfect dark matter fluid} \label{sec3}

Let us consider a general model for dark matter described by an non-conserved energy-momentum tensor for an imperfect pressureless fluid. Thus we can write it as \cite{Pimentel:2016jlm},

\begin{eqnarray}\label{s3.1}
T_{\mu \nu} = T_{\mu \nu}^{pf} + T_{\mu \nu}^{vis} + T_{\mu \nu}^{h},
\end{eqnarray}

with a perfect fluid contribution,

\begin{eqnarray}\label{s3.2}
T_{\mu \nu}^{pf} = \rho\, u_{\mu}u_{\nu},
\end{eqnarray}
a viscous term
\begin{eqnarray}\label{s3.3}
T_{\mu \nu}^{vis} = - \xi \, \Theta \, h_{\mu \nu} - 2 \, \eta \, \sigma_{\mu \nu},
\end{eqnarray}
and a heat-flow contribution
\begin{eqnarray}\label{s3.4}
T_{\mu \nu}^{h} = q_{\mu}u_{\nu} + q_{\nu}u_{\mu},
\end{eqnarray}
being,

\begin{eqnarray}\label{s3.5}
\sigma_{\mu \nu} = \frac{1}{2} \, \left( h^{\alpha}_{\,\, \mu} \, \nabla_{\alpha} u_{\nu} + h^{\alpha}_{\,\, \nu} \, \nabla_{\alpha} u_{\mu} \right) - \frac{1}{3} \, \Theta \, h_{\mu  \nu},
\end{eqnarray}
with
\begin{eqnarray}\label{s3.6}
h_{\mu \nu} = g_{\mu \nu} + u_{\mu}u_{\nu},
\end{eqnarray}
and
\begin{eqnarray}\label{s3.7}
\Theta = \nabla_{\alpha} u^{\alpha}.
\end{eqnarray}

Here $q_{\mu}$ is a general energy current, $\xi$ is the bulk viscosity parameter and $\eta$ is the shear viscosity parameter. Finally, we will consider that, due to a possible interaction between dark matter and another species (like dark energy), this energy-momentum tensor is not conserved, i.e.
\begin{eqnarray}\label{s3.8}
\nabla_{\mu} {T}^{\mu}_{\,\,\nu} = Q_{\nu},
\end{eqnarray}
Since we are considering only linear perturbations, we can analyze each contribution individually.

\subsection{Dark matter with bulk and shear viscosity} \label{sec3.1}

In this subsection we consider only the viscosity term so that $T_{\mu \nu} =  T_{\mu \nu}^{pf}+T_{\mu \nu}^{vis}$ and assume it is conserved. We consider a general perturbation in the bulk and shear viscosities,

\begin{eqnarray}\label{s3.1.1}
\xi(\tau,\bf{x}) = \xi_0(\tau) + \delta \xi (\tau,\bf{x}),
\end{eqnarray}
\begin{eqnarray}\label{s3.1.2}
\eta(\tau,\bf{x}) = \eta_0(\tau) + \delta \eta (\tau,\bf{x}).
\end{eqnarray}

These perturbations can always be related to the matter density contrast in the following form

\begin{eqnarray}\label{s3.1.3}
\delta \xi \equiv \frac{a \rho}{{\cal H}} \, \xi_p \, \delta,
\end{eqnarray}
\begin{eqnarray}\label{s3.1.4}
\delta \eta \equiv \frac{a \rho}{{\cal H}} \, \eta_p \, \delta,
\end{eqnarray}
where $\xi_p$ and $\eta_p$ are arbitrary  dimensionless functions of time and scale, and  the prefactors $a\rho/{\cal H}$ are introduced for convenience. Then the conservation equations  take the form of (\ref{s2.1}-\ref{s2.2}) with,

\begin{align}\label{s3.1.5}
&c_{11} = 1-6 \bar{\xi}, \,\,\,\, c_{12} = 9 (\xi_p - \bar{\xi}), \,\,\,\, c_{13} = 9 \bar{\xi}, \nonumber\\
&c_{21} = \frac{1}{1-3\bar{\xi}} \, \left[ 1-27\bar{\xi}^2+6\bar{\xi}-3\,\frac{\bar{\xi}'}{{\cal H}} + \left(\frac{4}{3} \bar{\eta} + \bar{\xi} \right) \, \frac{k^2}{{\cal H}^2} \right], \nonumber\\
&c_{22} = - \frac{3 \xi_p}{1-3\bar{\xi}} \, \frac{k^2}{{\cal H}^2}, \,\,\,\, c_{23} = \frac{1}{1-3\bar{\xi}}, \,\,\,\, c_{14} = c_{24} = 0,
\end{align}
being,
\begin{eqnarray}\label{s3.1.6}
\bar{\xi} = \frac{{\cal H} \xi_0}{a \rho}, \,\,\,\, \bar{\eta} = \frac{{\cal H} \eta_0}{a \rho}.
\end{eqnarray}
Notice that in order to avoid large modifications, in the sub-Hubble regime, $\bar{\xi}, \xi_p, \bar{\eta}\ll 1$. In this situation,
\begin{align} \label{s3.1.7}
&c_{21} = 1+\left(\frac{4}{3} \bar{\eta} + \bar{\xi} \right) \, \frac{k^2}{{\cal H}^2}, \,\,\, c_{22} = - 3 \xi_p \, \frac{k^2}{{\cal H}^2}, \nonumber\\
&c_{11} = c_{23} = 1, \,\,\, c_{12} = c_{13} = c_{14} = c_{24} = 0,
\end{align}

using these expressions and relations (\ref{s2.4}-\ref{s2.8}) we obtain,

\begin{align} \label{s3.1.8}
\mu_m = \mu + \frac{2 \xi_p}{\Omega_m (a)} \, \frac{k^2}{{\cal H}^2},
\end{align}
\begin{align} \label{s3.1.9}
\mu_d = 1 + \left(\frac{4}{3} \bar{\eta} + \bar{\xi} \right) \, \frac{k^2}{{\cal H}^2},
\end{align}
\begin{align} \label{s3.1.10}
\mu_{\theta} = 1.
\end{align}
%

\subsection{Dark matter with heat flux} \label{sec3.2}

Now we consider only the heat flux contribution so that  $T_{\mu \nu} = T_{\mu \nu}^{pf}+T_{\mu \nu}^{h}$ and assume it is conserved. As we are considering an isotropic background, the perturbed heat flux is in general,

\begin{eqnarray}\label{s3.2.1}
q_\mu=\left[q_0(\tau)+\delta q_0(\tau, {\bf x}),\delta q_i(\tau, \bf{x})\right],
\end{eqnarray}
Again we can always relate these perturbations to the dark matter perturbations  in the following way,

\begin{eqnarray}\label{s3.2.2}
\delta q_0 \equiv q_d \, a \rho \, \delta,
\end{eqnarray}
\begin{eqnarray}\label{s3.2.3}
\delta q_i \equiv q_v \, a \rho \, v_i,
\end{eqnarray}
where $q_d$ and $q_v$ are arbitrary dimensionless functions of time and scale. The conservation equations take also the form of (\ref{s2.1}-\ref{s2.2}) with,

\begin{align} \label{s3.2.4}
&c_{11} = \frac{1}{1-2q_d} \, [1+q_v-\bar{q}_0], \nonumber\\
&c_{12} = \frac{2}{1-2q_d} \, \left[\frac{q_d'}{{\cal H}} + \left( 2+\frac{(a \rho)'}{{\cal H} a \rho} \right) q_d \right. \nonumber\\
&\left. - \frac{\bar{q}_0'}{{\cal H}} - \left( 2+\frac{(a \rho)'}{{\cal H} a \rho} \right) \bar{q}_0 \right], \nonumber\\
&c_{13} = -\frac{2}{1-2q_d} \left[ \frac{\bar{q}_0'}{{\cal H}} + \left( 2+\frac{(a \rho)'}{{\cal H} a \rho} \right) \bar{q}_0 \right], \nonumber\\
&c_{21} = \frac{1}{1+q_v-q_0} \left[ 1 + \frac{q_v'}{{\cal H}} + \left( 3+\frac{(a \rho)'}{{\cal H} a \rho} \right) q_v \right. \nonumber\\
&\left. + \frac{\bar{q}_0'}{{\cal H}} + \left( 1+\frac{(a \rho)'}{{\cal H} a \rho} \right) \bar{q}_0 \right], \nonumber\\
&c_{23} = \frac{1}{1+q_v-q_0} \left[ 1-2 \bar{q}_0 \right], \nonumber\\
&c_{14} = c_{22} = c_{24} = 0,
\end{align}
being $\bar{q}_0 =q_0 / a \rho$. Then we can use equations (\ref{s2.4}-\ref{s2.8}) and obtain $\mu_m$, $\mu_d$ and $\mu_{\theta}$ that are in general different from one. They are lengthy
expressions which we do not show explicitly.

\subsection{Non-conserved dark matter} \label{sec3.3}

Finally, we consider that the energy-momentum tensor is not conserved in general. This may be due to an interaction between dark matter and dark energy. We perturb the $Q_{\mu}$ four-vector considering an isotropic background,

\begin{eqnarray}\label{s3.3.1}
Q_\mu=\left[Q_0(\tau)+\delta Q_0(\tau, {\bf x}),\delta Q_i(\tau, \bf{x})\right],
\end{eqnarray}
We relate these perturbations with the dark matter perturbations in the following way,

\begin{eqnarray}\label{s3.3.2}
\delta Q_0 \equiv \nu_0 \, \rho {\cal H} \, \delta,
\end{eqnarray}
\begin{eqnarray}\label{s3.3.3}
\delta Q_i \equiv \nu_v \, \rho {\cal H} \, v_i,
\end{eqnarray}
being $\nu_0$ and $\nu_v$ arbitrary dimensionless functions of time and scale.  The $c_{ij}$ coefficients of equations (\ref{s2.1}-\ref{s2.2}) are,

\begin{align} \label{s3.3.4}
&c_{12} = x - \nu_0, \,\,\,\,\, c_{21} = 1 - x - \nu_v, \nonumber\\
&c_{11} = c_{23} = 1, \,\,\, c_{13} = c_{14} = c_{22} = c_{24} = 0,
\end{align}
where $x = Q_0 / {\cal H} \rho$. If we use equations (\ref{s2.4}-\ref{s2.8}) we obtain,

\begin{align} \label{s3.3.5}
\mu_m = \mu + \frac{2}{3 \, \Omega_m (a)} \, &\left[ \nu_0 \, \nu_v - x \, \nu_v - (1-x) \, (\nu_0-x) \right. \nonumber\\ 
&\left. + \frac{({\cal H} x)'}{{\cal H}^2} - \frac{({\cal H} \nu_0)'}{{\cal H}^2} \right],
\end{align}
\begin{align} \label{s3.3.6}
\mu_d = 1 + \nu_0 - \nu_v - 2\,x,
\end{align}
\begin{align} \label{s3.3.7}
\mu_{\theta} = 1 - \frac{{\cal H} \, \delta}{\delta'} \, (x-\nu_0).
\end{align}

To summarize what  we have shown in the last subsections, a general model for an imperfect non-conserved fluid in a modified gravity scenario  can be described with parameters $(\mu, \gamma, \mu_m, \mu_d, \mu_{\theta})$. The corresponding  modified gravity and modified dark matter equations are,

\begin{eqnarray}\label{s3.3.8}
k^2 \, \Phi = -\frac{3}{2} \, {\cal H}^2 \, \Omega_m (a) \mu \, \gamma \, \delta,
\end{eqnarray}
\begin{eqnarray}\label{s3.3.9}
k^2 \, \Psi = -\frac{3}{2} \, {\cal H}^2 \, \Omega_m (a) \mu \, \delta,
\end{eqnarray}
\begin{eqnarray}\label{s3.3.10}
\delta'' + {\cal H} \, \mu_d \, \delta' - \frac{3}{2} \, {\cal H}^2 \, \Omega_m (a) \, \mu_m \, \delta = 0,
\end{eqnarray}
\begin{eqnarray}\label{s3.3.11}
\theta = - \mu_{\theta} \, \delta'.
\end{eqnarray}
%

\section{Growth function parametrization} \label{sec4}

As we are interested in obtaining how the growth function changes due to the modified growth equation (\ref{s3.3.10}), in this section we will calculate useful analytical approximations in terms of $\mu_m$ and $\mu_d$ parameters. We will use these analytical approximations to obtain forecast results in a simple way using the Fisher matrix. The growth function is defined as,

\begin{eqnarray}\label{s4.1}
f(a) \equiv \frac{d \, \ln \delta}{dN},
\end{eqnarray}
being $N = \ln a$. We will consider two cases: first of all the simplest case in which $\mu_m$ and $\mu_d$ are just constants, and then the case in which $|1-\mu_m(a,k)| \ll 1$ and $|1-\mu_d(a,k)| \ll 1$.

We start with the growth equation (\ref{s3.3.10}) and we do the following variable change,

\begin{eqnarray}\label{s4.2}
\delta' = {\cal H} \, \delta \, f,
\end{eqnarray}
so that
\begin{eqnarray}\label{s4.3}
\delta'' = {\cal H}^2 \, \delta \, \left[ \dot{f} + f^2 + \frac{\dot{{\cal H}}}{{\cal H}} \,f \right],
\end{eqnarray}
where dot denotes derivative respect to $N$. Using (\ref{s4.2}-\ref{s4.3}) in (\ref{s3.3.10}) and ${\cal H} = a H$ we obtain,

\begin{eqnarray}\label{s4.4}
\dot{f} + f^2 + \left[ 1 + \mu_d + \frac{\dot{H}}{H} \right] \,f = \frac{3}{2} \, \Omega_m (a) \, \mu_m .
\end{eqnarray}

For the standard case $(\mu_m, \mu_d)=(1,1)$ we have the known analytical fitting function  \cite{Linder:2007hg},

\begin{eqnarray}\label{s4.5}
f(a) = \Omega_m^{\hat\gamma} (a),
\end{eqnarray}
where $\hat\gamma = 0.545$ is the growth index \cite{Linder:2007hg}. Also, when $\mu_m = const \neq 1$ we found a good analytical fitting function \cite{Resco:2017jky},

\begin{eqnarray}\label{s4.5.1}
f(a) = \frac{1}{4} \left[ \sqrt{1+24\mu_m} - 1 \right] \, \Omega_m^{\hat\gamma} (a).
\end{eqnarray}

\begin{figure} 
  	\includegraphics[width=0.495\textwidth]{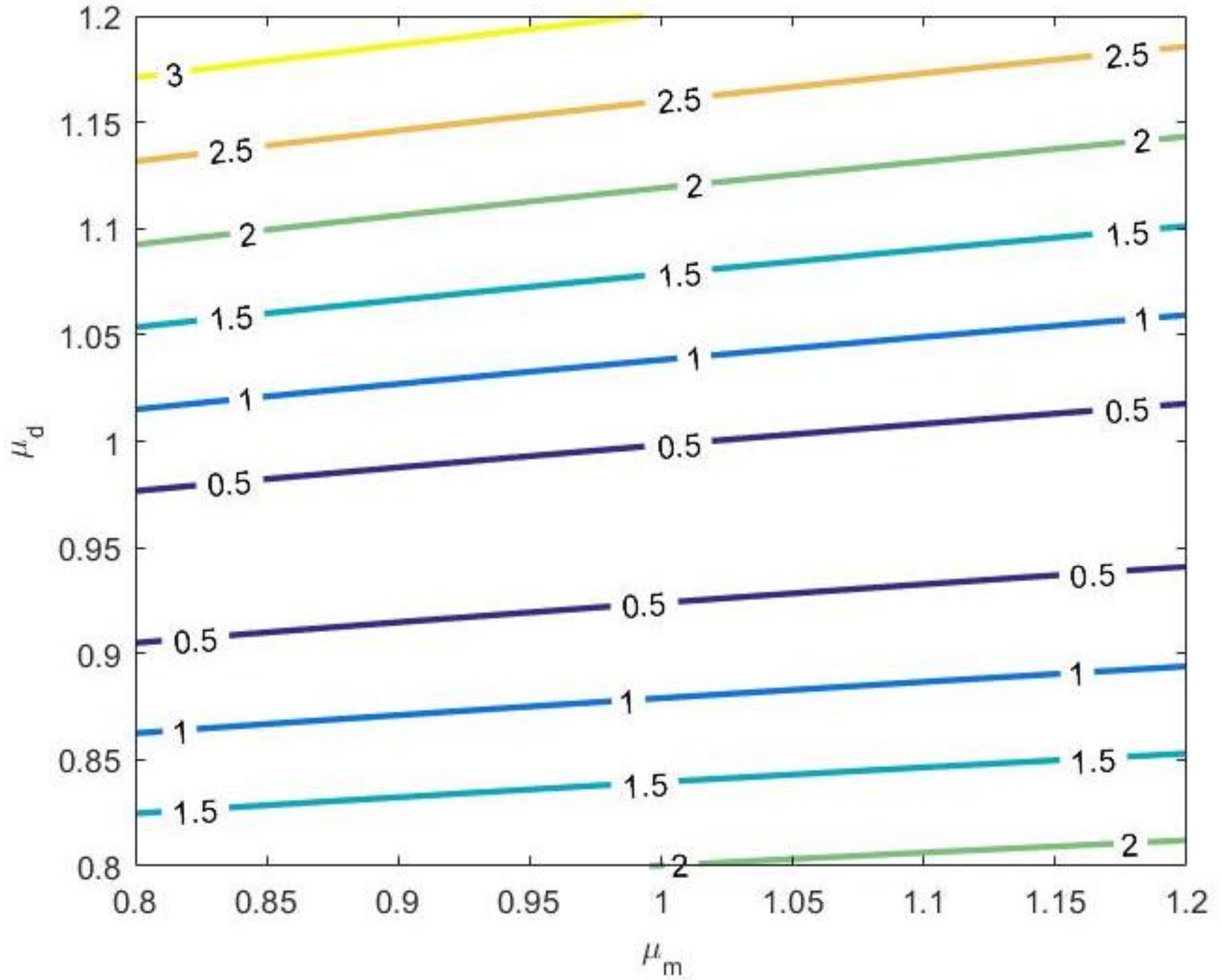}
		\caption{Errors (\%) of the analytic expression (\ref{s4.10}) with respect to the numerical solution of $f(a)$ for constant $\mu_m$ and $\mu_d$.}
  \label{Figure_1}
\end{figure}

We want to include the $\mu_d$ effect so we proceed as in \cite{Resco:2017jky}. Let us  assume a  solution of the form,

\begin{eqnarray}\label{s4.6}
f(a) = \zeta (\mu_m, \mu_d) \, \Omega_m^{\hat\gamma} (a),
\end{eqnarray}
in a $\mathrm{\Lambda CDM}$ background so that
\begin{eqnarray}\label{s4.7}
\frac{\dot{H}}{H} = - \frac{3}{2} \, \Omega_m (a),
\end{eqnarray}
with
\begin{eqnarray}\label{s4.8}
\dot{\Omega}_m (a) = - 3 \, \Omega_m (a) \, (1-\Omega_m (a)),
\end{eqnarray}
using (\ref{s4.6}-\ref{s4.8}) in (\ref{s4.4}) and considering the approximation $\Omega_m (a) \simeq 1$ we obtain,
\begin{eqnarray}\label{s4.9}
\zeta^2 - \left[ \frac{1}{2} - \mu_d \right] \, \zeta - \frac{3}{2} \, \mu_m = 0,
\end{eqnarray}
Thus we can obtain $\zeta (\mu_m, \mu_d)$ and extend the result in (\ref{s4.5.1}) as,

\begin{eqnarray}\label{s4.10}
f(a) = \frac{1}{4} \left[ \sqrt{(1 - 2 \mu_d)^2+24\mu_m} + 1 - 2 \mu_d \right] \, \Omega_m^{\hat\gamma} (a). \nonumber\\
\end{eqnarray}
Let us know check the accuracy of this analytic approximation with respect to the numerical solution. We plot the errors in Fig. \ref{Figure_1}. As we can see, if the modifications of $\mu_m$ and $\mu_d$ are below the $10 \, \%$, the analytic approximation has errors of order $1 \, \%$.

Now we analyze the case $|1-\mu_m(a,k)| \ll 1$ and $|1-\mu_d(a,k)| \ll 1$. Proceeding as in \cite{Resco:2017jky}, we assume a solution of the form,

\begin{eqnarray}\label{s4.11}
f(a,k) = \left[1 + \epsilon(a,k) \right] \, \Omega_m^{\hat\gamma} (a),
\end{eqnarray}
with $\epsilon \ll 1$. We use (\ref{s4.11}) into (\ref{s4.4}), keeping linear terms in $1-\mu_m$, $1-\mu_d$ and $\epsilon$. Considering the approximation $\Omega_m (a) \simeq 1$ we obtain,

\begin{eqnarray}\label{s4.12}
\dot{\epsilon} + \frac{5}{2} \, \epsilon = \left(1 - \mu_d \right) - \frac{3}{2} \, \left(1 - \mu_m \right),
\end{eqnarray}
Solving this last equation and changing to the variable $z$, we obtain the correction to the growth function $\epsilon$ as,

\begin{widetext}

\begin{eqnarray}\label{s4.13}
\epsilon(z,k) = \left( 1+z \right)^{5/2} \, \int_z^{z_{mat}} \, \left[ \left(1 - \mu_d (z',k) \right) - \frac{3}{2} \, \left(1 - \mu_m (z',k) \right) \right] \, \left( 1+z' \right)^{-7/2} \, dz',
\end{eqnarray}

\end{widetext}
where $z_{mat}$ is the redshift in the matter dominated era for which $\mu_m(z_{mat},k) \simeq \mu_d(z_{mat},k) \simeq  1$ that we take as $z_{mat} \simeq 10$ (the results are not very sensitive to the specific value chosen). 

In order to test this approximation we consider the following expressions for $\mu_m (z)$ and $\mu_d (z)$,

\begin{eqnarray}\label{s4.14}
\mu_m (z) = 1 + (\mu_m^0 - 1) \, \frac{1-\Omega_m (z)}{1-\Omega_m},
\end{eqnarray}
and
\begin{eqnarray}\label{s4.15}
\mu_d (z) = 1 + \frac{(\mu_d^0 - 1)}{1+z}.
\end{eqnarray}
\begin{figure} 
  	\includegraphics[width=0.495\textwidth]{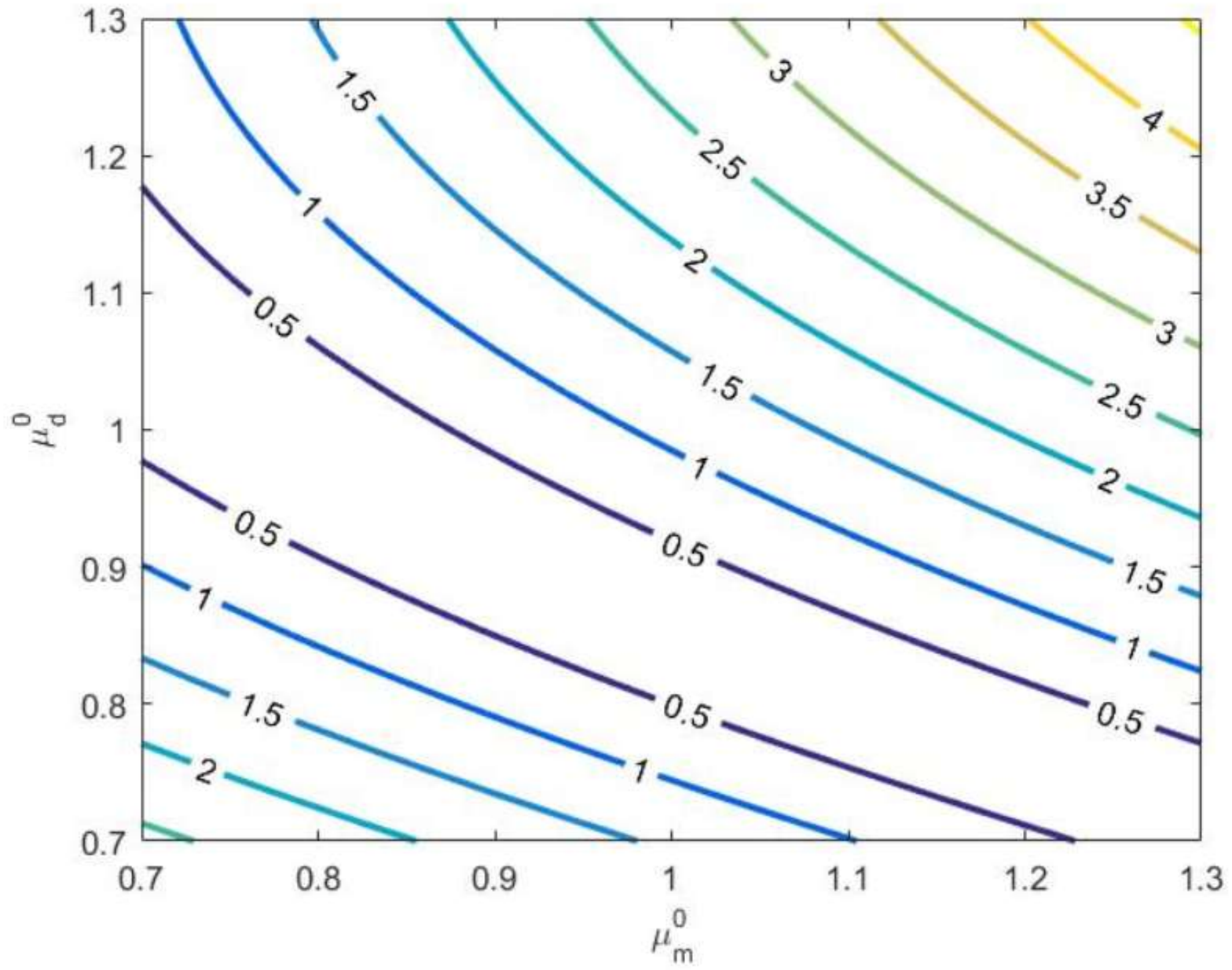}
		\caption{Errors (\%) of the analytic expression (\ref{s4.13}) with respect to the numerical solution of $f(z)$ for $\mu_m (z)$ and $\mu_d (z)$ following expressions (\ref{s4.14}-\ref{s4.15}).}
  \label{Figure_1a}
\end{figure}

We plot the errors for the growth function from the approximation \eqref{s4.13} with respect to the numerical solution in Fig. \ref{Figure_1a}. As in the previous case, the analytic approximation has errors of order $1 \, \%$ for modifications of $\mu_m$ and $\mu_d$ of order $10 \, \%$. 

In addition, we can do an analytical check of expression (\ref{s4.11}) for the case in which $\mu_m$ and $\mu_d$ are constants. Using,

\begin{eqnarray}\label{s4.16}
\left( 1+z \right)^{5/2} \, \int_z^{z_{mat}} \, \left( 1+z' \right)^{-7/2} \, dz' \simeq \frac{2}{5},
\end{eqnarray}

for $z \ll z_{mat}$, we obtain,

\begin{eqnarray}\label{s4.17}
f(a) = \left[ \frac{4}{5} - \frac{2}{5} \mu_d + \frac{3}{5} \mu_m  \right] \, \Omega_m^{\hat\gamma} (a).
\end{eqnarray}

We also recover this result if we apply the aproximation $|1-\mu_m| \ll 1$ and $|1-\mu_d| \ll 1$ in equation (\ref{s4.10}).

\section{Observable effects in galaxy and weak lensing power spectra} \label{sec5}

The next era of precision cosmology will be driven by galaxy surveys, so it is interesting to study how sensitive the galaxy observables are to the effective parameters introduced before. In this section we are going to calculate the galaxy and weak lensing power spectra: $P_{gg}$, $P_{u u}$, $P_{g u}$ and $P_{\kappa\kappa}$ \cite{White:2008jy, McDonald:2008sh, Lemos:2017arq, Guzik:2009cm, Kilbinger:2014cea} in the presence of a modified cosmology and a modified dark matter fluid (\ref{s3.3.7}-\ref{s3.3.10}), where $u$ is the line of sight velocity that we define below.

First of all, we calculate the $P_{gg}$, $P_{u u}$, $P_{g u}$ power spectra. We need to take into account that we measure the galaxy  density contrast in  redshift space so that \cite{Bernardeau:2001qr},

\begin{eqnarray}\label{s5.1}
\delta^s_g = \delta^r_g - \hat{\mu}^2 \, \frac{\theta_g}{{\cal H}},
\end{eqnarray}
where $\delta^s_g$ and $\delta^r_g$ are the density contrast of galaxies in  redshift space and real space respectively $\hat{\mu}$ is the cosine of the angle between $\vec{k}$ and the line of sight, and $\theta_g = i k_i v_g^i$. We consider only a bias $b$ between the density contrast of galaxies and dark matter so that
\begin{eqnarray}\label{s5.2}
\delta^r_g = b \, \delta, \,\,\,\, \theta_g = \theta,
\end{eqnarray}
Using equations (\ref{s5.2}), (\ref{s3.3.10}) and (\ref{s4.2}) in equation (\ref{s5.1}) we find,
\begin{eqnarray}\label{s5.3}
\delta^s_g = \left( 1 + \mu_{\theta} \, \beta \, \hat{\mu}^2 \right) \, b \, \delta,
\end{eqnarray}
where we have defined $\beta \equiv f/b$. Also, the line of sight velocity is defined as,
\begin{equation}\label{s5.4}
u (k, \hat{\mu}) = i \, \hat{\mu} \, \frac{\theta_g}{k},
\end{equation}

Thus, taking into account that $\theta_g = \theta = - \mu_{\theta} \, {\cal H} \, f \, \delta$,  we obtain,

\begin{eqnarray}\label{s5.5}
u = - \frac{i \, \mu_{\theta} \, H \, f \, \hat{\mu}}{(1+z) \, k} \, \delta.
\end{eqnarray}
\\
Using the expressions for $\delta^s_g$ and $u$ we can calculate the power spectra: $P_{g g} = \langle \delta^{s}_g \delta^{s \,\, *}_g \rangle$, $P_{u u} = \langle u u^* \rangle$ and $P_{g u} = \langle \delta^{s}_g u^* \rangle$,

\begin{eqnarray}\label{s5.6}
P_{gg} (z, k_{r}, \hat{\mu}_{r}) = \frac{D_{A \,\, r}^2 \, H}{D_{A}^2 \, H_r} \, \left( 1 + \mu_{\theta} \, \beta \, \hat{\mu}^2 \right)^2 \, D^2 \, b^2 \, P(k), \nonumber\\ 
\end{eqnarray}
\begin{equation}\label{s5.7}
P_{uu} (z, k_r, \hat{\mu}_r) = \frac{D_{A \,\, r}^2 \, H}{D_{A}^2 \, H_r} \, \left( \frac{\mu_{\theta} \, H \, f \, \hat{\mu}}{(1+z) \, k} \right)^2 \, D^2 \, P(k),
\end{equation}
{\small
\begin{align}\label{s5.8}
P_{gu} (z, k_r, \hat{\mu}_r) = \frac{D_{A \,\, r}^2 \, H}{D_{A}^2 \, H_r} \, \frac{i \, \mu_{\theta} \, H \, f \, \hat{\mu}}{(1+z) \, k} \, \left( 1 + \mu_{\theta} \, \beta \, \hat{\mu}^2 \right) \, D^2 \, b \, P(k),
\end{align}
}
where we have taken into account the Alcock-Paczynski effect \cite{Alcock:1979mp} being,

\begin{equation}\label{s5.9}
k=Q \, k_{r},
\end{equation}
\begin{equation}\label{s5.10}
\hat{\mu}=\frac{H \, \hat{\mu}_{r}}{H_{r} \, Q},
\end{equation}
\begin{equation}\label{s5.11}
Q=\frac{\sqrt{H^{2} \, \chi^{2} \, \hat{\mu}^{2}_{r}-H_{r}^{2} \, \chi_{r}^{2} \, (\hat{\mu}^{2}_{r}-1)}}{H_{r} \, \chi},
\end{equation}
where the subindex $r$ denotes that the corresponding quantity is evaluated in the
fiducial cosmology, $\chi$ is the comoving radial distance,
\begin{equation}\label{s5.12}
\chi(z)=\int_{0}^{z} \frac{dz'}{H(z')},
\end{equation}
and $D_A$ is the angular distance. In a flat Universe $D_A (z) = (1+z)^{-1} \, \chi(z)$. $P(k)$ is the dark matter power spectrum today and $D(z) = \delta(z) / \delta(0)$,

\begin{equation}\label{s5.13}
D(z) = \textrm{exp} \left[ \int_{0}^{N(z)} \, f(N') dN' \right],
\end{equation}
being $N(z) = - \ln (1+z)$. As we can see from (\ref{s4.10}), $D(z) = D(z, \mu_m, \mu_d)$ and $f(z) = f(z, \mu_m, \mu_d)$. Finally, the convergence power spectrum is affected by the combination $k^2 (\Psi + \Phi)$ so we obtain the result,

\begin{widetext}

\begin{equation}\label{s5.14}
P_{\kappa\kappa \,\, i j} (\ell) = \frac{9 \, H_0^4 \, \Omega_m^2}{4} \, \int_0^{\infty} \, dz \, \frac{(1+z)^2}{H(z)} \, g_i(z) \, g_j(z) \, \frac{\mu^2 \, (1+\gamma)^2}{4} \, D(z, \mu_m, \mu_d)^2 \, P \left( \frac{\ell}{\chi (z)} \right),
\end{equation}

\end{widetext}

where $g_i(z)$ is the window function,

\begin{equation}\label{s5.15}
g_i(z) = \int_z^{\infty} \, \left( 1-\frac{\chi (z)}{\chi(z')} \right) \, n_i(z') \, dz',
\end{equation}
being $n_i (z)$ the galaxy density, normalized to one, in the redshift bin $i$,

\begin{equation}\label{s5.16}
n_i (z) \propto \int_{\bar{z}_{i-1}}^{\bar{z}_i} \, n(z') e^{\frac{(z'-z)^2}{2 \sigma_i^2}} \, dz',
\end{equation}
where the redshift error is $\sigma_i = \delta z (1+z_i)$, $\bar{z}_i$ is the upper limit of the $i$-bin and $n(z)$ is,

\begin{equation}\label{s5.17}
n(z) = \frac{3}{2 z_p^3} \, z^2 \, e^{-(z/z_p)^{3/2}},
\end{equation}
being $z_p = z_{mean}/\sqrt{2}$ and $z_{mean}$ the survey mean redshift. As we can see from the power spectra observables, $P_{gg}$, $P_{uu}$ and $P_{gu}$ are sensitive to $(\mu_{\theta}, \zeta)$; and $P_{\kappa\kappa}$ is sensitive to $(\Sigma, \zeta)$ where,

\begin{equation}\label{s5.18}
\zeta = \frac{1}{4} \left[ \sqrt{(1 - 2 \mu_d)^2+24\mu_m} + 1 - 2 \mu_d \right],
\end{equation}
and
\begin{equation}\label{s5.19}
\Sigma = \frac{\mu \, (1+\gamma)}{2},
\end{equation}

Thus, as shown before, we need $(\mu, \gamma, \mu_m, \mu_d, \mu_{\theta})$ to describe, in a model independent way, a general modification of Einstein equations and conservation equations, however, using galaxy power spectra, as we can see in equations (\ref{s5.6}-\ref{s5.8}) and (\ref{s5.14}), we are only sensitive to the following combinations: $D \, b$, $\mu_{\theta} \, D \, f$, $\Sigma \, D$ and the Hubble parameter $H$. Assuming that the galaxy bias is fixed by a bias model, the observables are able to constrain
the three combinations $(\Sigma, \zeta, \mu_{\theta})$. Therefore, parameters $(\mu, \gamma)$ and $(\mu_{m}, \mu_{d})$ are degenerated.

\begin{table}[htbp]
\begin{tabular}{|c|c|c|c|}
\hline
$\Sigma$ & $\zeta$ & $\mu_{\theta}$ & Underlying theory  \\
\hline \hline 
 1 & 1 & 1 & $\mathrm{\Lambda CDM}$  \\ \hline
 $\neq 1$ & 1 & 1 & $MG \,\,\, \gamma \neq 1$  \\ \hline
 1 & $\neq 1$ & 1 & $BSV$  \\ \hline
 1 & 1 & $\neq 1$ & -  \\ \hline
 1 & $\neq 1$ & $\neq 1$ & $BSV + HNC$   \\ \hline
 $\neq 1$ & 1 & $\neq 1$ & -  \\ \hline
 $\neq 1$ & $\neq 1$ & 1 & $MG + BSV$  \\ \hline
 $\neq 1$ & $\neq 1$ & $\,\,\neq 1$ & $MG + BSV + HNC\,\,$   \\ \hline
\end{tabular}
\caption{$MG$ denotes modified gravity, $BSV$  bulk and shear viscosity, and $HNC$ refers to heat flux and non-conserved fluid. If the underlying theory is $-$ means that this possible theory has a fine tuning parameter values.}
\label{Table1}
\end{table}

Any deviation from one of any of the three mentioned parameters will imply a modification of General Relativity or a modification of the perfect-fluid description of dark matter. We summarize in Table \ref{Table1} the different combinations of parameters and the  compatible underlying theories. Thus, for example, we see that a 
detection of $\Sigma\neq 1$ with $\zeta= \mu_{\theta}=1$ is only compatible with modified gravity. On the other hand, $\zeta\neq 1$ with  $\Sigma=\mu_{\theta}= 1$ can only be produced by modified dark matter. Similarly, $\zeta\neq 1$ and $\mu_{\theta}\neq 1$ with  $\Sigma= 1$ cannot be generated by modified gravity. On the contrary, we see that different from one measurements of the three parameters will not allow to distinguish whether the underlying theory is a modification of gravity or an imperfect dark matter.

\section{Fisher forecast from clustering, velocity and weak lensing power spectra} \label{sec6}

Once we have some useful parametrizations of the growth function and we have seen how galaxy power spectra are modified, we want to perform a Fisher analysis to forecast the capability of future surveys like Euclid to constrain the dark matter phenomenological parametrization. Therefore, let  us summarize the Fisher analysis for the $P_{gg}$, $P_{uu}$, $P_{gu}$ and $P_{\kappa\kappa}$ power spectra. We are interested in computing the Fisher matrix in each redshift bin. Thus, the Fisher matrices for $P_{gg}$, $P_{uu}$ and $P_{gu}$ are \cite{Howlett:2017asw},

\begin{widetext}

\begin{equation}\label{s6.1}
F_{\alpha \beta} (z_a) = \frac{V(z_a)}{8 \pi^2} \, \int_{-1}^1 d\hat{\mu} \, \int_{k_{\text{min}}}^{k_{\text{max}}} dk \, k^2 \, \left.\frac{\partial \Sigma_{i j}}{\partial p_{\alpha}}\right|_{r} \, \Sigma_{j m}^{-1} \, \left.\frac{\partial \Sigma_{m n}}{\partial p_{\beta}}\right|_{r} \, \Sigma_{n i}^{-1},
\end{equation}

\end{widetext}
where $V(z_a)$ is the comoving volume of the redshift bin $a$,

\begin{equation}\label{s6.2}
V(z_a) = \frac{4 \pi \, f_{sky}}{3} \, \left[ \chi^3 (z_a+\Delta z_a / 2) - \chi^3 (z_a-\Delta z_a / 2) \right],
\end{equation}
being $f_{sky}$ the fraction of the sky and $\Delta z_a$ the width of the bin $a$. The matrix $\Sigma_{i j}$ is,

\begin{equation}\label{s6.3}
\bf{\Sigma} = \left( \begin{array}{cc}
 P_{gg} + \bar{n}_g^{-1} & P_{gu}  \\
 P_{ug} & P_{uu} + \bar{n}_u^{-1} \, \sigma_u^2  \end{array} \right),
\end{equation}
where $\bar{n}_g$ is the galaxy density in bin $a$ and $\bar{n}_u$ is the galaxy density for the velocity field in bin $a$. In general $\bar{n}_g > \bar{n}_u$. $\sigma_u$ is the velocity noise,
\begin{equation}\label{s6.4}
\sigma_u^2 = \sigma_*^2 + \epsilon \, z,
\end{equation}
being $\sigma_* = 10^{-3}$ $(c=1)$ and $\epsilon=0.2$ the fractional error \cite{Koda:2013eya}. We can obtain the Fisher matrix only for $P_{gg}$, $P_{uu}$ or $P_{gu}$ using expression (\ref{s6.1}),

\begin{widetext}

\begin{equation}\label{s6.5}
F_{\alpha \beta}^{gg} (z_a) = \frac{V(z_a)}{8 \pi^2} \, \int_{-1}^1 d\hat{\mu} \, \int_{k_{\text{min}}}^{k_{\text{max}}} dk \, k^2 \, \left.\frac{\partial \ln P_{gg}}{\partial p_{\alpha}}\right|_{r} \, \left.\frac{\partial \ln P_{gg}}{\partial p_{\beta}}\right|_{r} \, \left[ \frac{\bar{n}_g P_{gg}}{1+\bar{n}_g P_{gg}} \right]^2,
\end{equation}
\begin{equation}\label{s6.6}
F_{\alpha \beta}^{uu} (z_a) = \frac{V(z_a)}{8 \pi^2} \, \int_{-1}^1 d\hat{\mu} \, \int_{k_{\text{min}}}^{k_{\text{max}}} dk \, k^2 \, \left.\frac{\partial \ln P_{uu}}{\partial p_{\alpha}}\right|_{r} \, \left.\frac{\partial \ln P_{uu}}{\partial p_{\beta}}\right|_{r} \, \left[ \frac{\sigma_u^{-2} \bar{n}_u P_{uu}}{1+\sigma_u^{-2} \bar{n}_u P_{uu}} \right]^2,
\end{equation}

\end{widetext}

\begin{widetext}

\begin{equation}\label{s6.7}
F_{\alpha \beta}^{gu} (z_a) = \frac{V(z_a)}{4 \pi^2} \, \int_{-1}^1 d\hat{\mu} \, \int_{k_{\text{min}}}^{k_{\text{max}}} dk \, k^2 \, \left.\frac{\partial \ln P_{gu}}{\partial p_{\alpha}}\right|_{r} \, \left.\frac{\partial \ln P_{gu}^{*}}{\partial p_{\beta}}\right|_{r} \, \left[ \frac{\sigma_u^{-2} \bar{n}_u \bar{n}_g P_{gu} P_{gu}^{*}}{\left( 1+\bar{n}_g P_{gg} \right)\left( 1+\sigma_u^{-2} \bar{n}_u P_{uu} \right) + \sigma_u^{-2} \bar{n}_u \bar{n}_g P_{gu} P_{gu}^{*}} \right].
\end{equation}

\end{widetext}

Finally, the Fisher matrix for the convergence power spectrum is,

\begin{widetext}
\begin{equation}\label{s6.8}
F_{\alpha \beta} = f_{sky} \, \sum_{\ell} \, \Delta \ln \ell \, \frac{(2 \ell + 1) \, \ell}{2} \, \left.\frac{\partial P_{\kappa\kappa \,\, i j}}{\partial p_{\alpha}}\right|_{r} \, C_{j m}^{-1} \, \left.\frac{\partial P_{\kappa\kappa \,\, m n}}{\partial p_{\beta}}\right|_{r} \, C_{n i}^{-1},
\end{equation}
\end{widetext}
where

\begin{align}\label{s6.9}
C_{i j} = P_{\kappa\kappa \,\, i j} + \frac{\gamma_{int}^2}{\hat{n}_i} \, \delta_{i j},
\end{align}
being $\gamma_{int} = 0.22$ the intrinsic ellipticity \cite{Hilbert:2016ylf}, $\hat{n}_{i}$ the galaxies per steradian in the $i$-th bin,

\begin{equation}\label{s6.10}
\hat{n}_{i}=n_{\theta} \, \frac{\int_{\bar{z}_{i-1}}^{\bar{z}_{i}} n(z) \, dz}{\int_{0}^{\infty} n(z) \, dz},
\end{equation}
where $n_{\theta}$ is the areal galaxy density and $n(z)$ follows equation (\ref{s5.17}). We sum in $\ell$ with $\Delta \ln \ell = 0.1$ from $\ell_{min} = 5$ to $\ell_{max} = \chi (z_{\alpha'}) \, k_{\text{max}}$ being $\alpha' = \mathrm{min}(\alpha, \beta)$. For this Fisher matrix we need to discretize in redshift bins the convergence power spectrum (\ref{s5.14}),

\begin{widetext}

\begin{equation}\label{s6.11}
P_{\kappa\kappa \,\, i j} (\ell) = \frac{9 H_0^4 \Omega_m^2}{4} \, \sum_a \Delta z_a \, \frac{(1+z_a)^2}{H_a} \, g_i (z_a) g_j (z_a) \, \frac{\mu_a^2 (1+\gamma_a)^2}{4} \, D_a^2 \, P \left( \frac{\ell}{\chi (z_a)} \right).
\end{equation}

\end{widetext}

The size of the Fisher matrix (\ref{s6.5}) will be $n_p \times n_a$ where $n_p$ is the number of parameters and $n_a$ is the total number of redshift bins. Further details on the computation of Fisher matrices can be found in \cite{FARO}.

The fiducial cosmology we consider is given by
the Planck values \cite{Aghanim:2018eyx} $\Omega_{c} \, h^{2}=0.121$, $\Omega_{b} \, h^{2}=0.0226$, $\Omega_{\nu} \, h^{2}=0.00064$, $n_{s}=0.96$, $h=0.68$, $H^{-1}_{0}=2997.9 \, \textrm{Mpc/h}$, $\Omega_{k}=0$ and $\sigma_{8}=0.82$ in the standard \textrm{$\Lambda$CDM} model. For this cosmology $E(z) \equiv H(z) / H_0$, reads
\begin{equation}\label{26}
E(z)=\sqrt{\Omega_{m} \, (1+z)^{3}+(1-\Omega_{m})}.
\end{equation}

For the fiducial cosmology we obtain the present matter power spectrum $P(k)$ from CLASS  \cite{Lesgourgues:2011re}. Finally, because we use \textrm{$\Lambda$CDM} for the fiducial model, $[\Sigma, \zeta, \mu_{\theta}]|_{r} = [1, 1, 1]$.

\subsection{Euclid survey} \label{sec6.2}

Euclid is a spectroscopy and photometric survey that will be able to cover 15000 $\mathrm{deg^2}$ up to redshift $z=2$ with a very high number of emission-line galaxies. In this subsection we would forecast the precision of Euclid in the measurements of the effective parameters from  $P_{gg}$ and $P_{\kappa\kappa}$ power spectra. The specifications of Euclid are \cite{Amendola:2013qna}:  fraction of the sky is $f_{sky} = 0.364$. The bias is,

\begin{equation}\label{s6.2.1}
b(z)=\sqrt{1+z},
\end{equation}
for $k_{\text{max}}$ values, we consider $k_{\text{max}}=k_{\text{max}}(z_{a})$ imposing that $\sigma^2(z_{a},\pi/2k_{\text{max}}(z_{a}))=0.35$ so that we only consider modes in the linear regime,
\begin{eqnarray}\label{s6.2.2}
\sigma^{2}(z,R) = D^2(z) \, \int{\frac{k'^{2}\, dk'}{2\pi^{2}} P(k') |\hat{W}(R,k')|^{2}} ,
\end{eqnarray}
We use a top-hat filter $\hat{W}(R,k)$, defined by
\begin{equation}\label{s6.2.3}
\hat{W}(R,k)=\frac{3}{k^{3}R^{3}} \, [\sin(kR)-kR\cos(kR)].
\end{equation}

For $k_{\text{min}}$ we take the value $k_{\text{min}} = 0.007$ $h$/Mpc \cite{Amendola:2013qna}.

\begin{widetext}

\begin{table}[htbp]
\begin{tabular}{|c|c|c|c|c|c|c|c|c|c|}
\hline
$z$ & $k_{\text{max}}$ & $\ell_{\text{max}}$ & $n \, \times \, 10^{-3}$ & $\Delta \mu_{\theta}^C / \mu_{\theta} (\%)$ & $\Delta \zeta^C / \zeta (\%)$ & $\Delta \Sigma^L / \Sigma (\%)$ & $\Delta \mu_{\theta}^T / \mu_{\theta} (\%)$ & $\Delta \zeta^T / \zeta (\%)$ & $\Delta \Sigma^T / \Sigma (\%)$  \\
\hline \hline 
0.6 & 0.195 & 300  & 3.56 & 3.03 & 2.61 &  2.90  &  3.00  &  2.58  &  0.88  \\ \hline
0.8 & 0.225 & 437  & 2.42 & 1.78 & 1.46 &  2.08  &  1.66  &  1.36  &  1.58  \\ \hline
1.0 & 0.260 & 597  & 1.81 & 1.25 & 0.97 &  3.60  &  1.18  &  0.92  &  2.96  \\ \hline
1.2 & 0.299 & 782  & 1.44 & 1.00 & 0.75 &  8.96  &  0.97  &  0.72  &  7.22  \\ \hline
1.4 & 0.343 & 994  & 0.99 & 0.96 & 0.68 &  30.0  &  0.95  &  0.67  &  23.7  \\ \hline
1.8 & 0.447 & 1510 & 0.33 & 0.84 & 0.56 &  340   &  0.83  &  0.56  &  249   \\ \hline
\end{tabular}
\caption{Redshift bins, $k_{\text{max}}$ in $\textrm{h/Mpc}$, $\ell_{\text{max}}$ values, galaxy densities in $(\textrm{h/Mpc})^{3}$ and relative errors for $\mu_{\theta}$, $\zeta$ and $\Sigma$ for Euclid survey. Super-index $C$ denotes clustering information, $L$ denotes lensing information and $T$ denotes clustering + lensing information.}
\label{Table3}
\end{table}

\end{widetext}

Errors in $z$ are $\delta z_{C} = 0.001$ for clustering. The values for redshift bins, galaxy densities and $k_{\text{max}}$ can be found in Table \ref{Table3}. Finally, for the convergence power spectra we take $z_{mean}=0.9$, $n_{\theta}=35$ galaxies per square arc minute with $\delta z_{L}=0.05$.

\begin{figure} 
  	\includegraphics[width=0.52\textwidth]{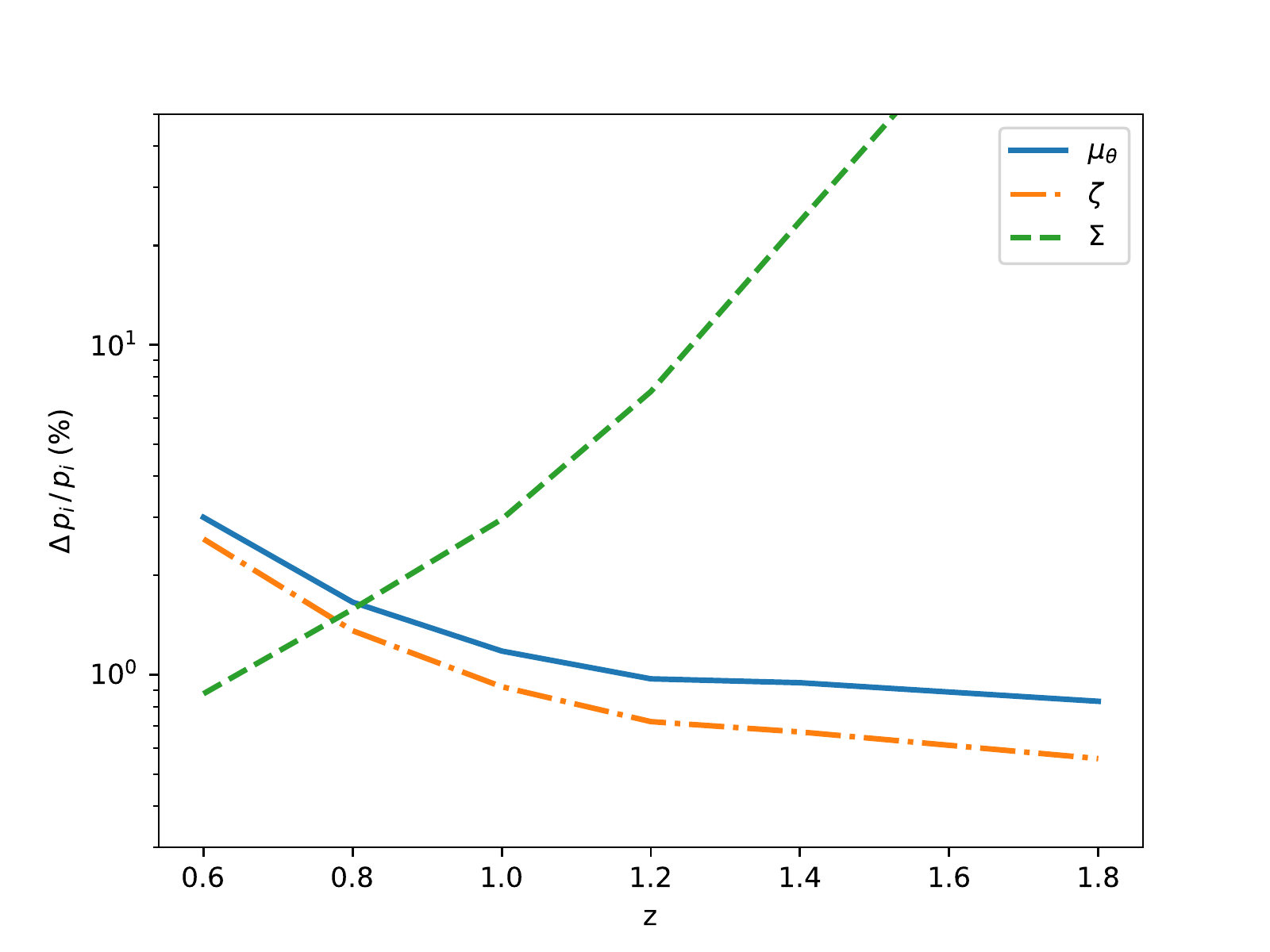}
		\caption{Forecasted errors (\%) for $\Sigma$, $\zeta$ and $\mu_{\theta}$ using clustering and lensing information for the  Euclid survey.}
  \label{Figure_2a}
\end{figure}

We will first consider the case in which the effective parameters are just constants. As independent parameters, we consider $(\mu_{\theta}, \zeta, E)$ for clustering and $(E, \Sigma)$ for lensing in each redshift bin. Then we combine clustering and lensing information in $(\mu_{\theta}, \zeta, \Sigma, E)$. We summarize the forecasted errors in Table \ref{Table3} and Fig. \ref{Figure_2a}. If we sum all the information in each bin  we find $\delta \mu_{\theta} / \mu_{\theta} = 0.43 \%$, $\delta \zeta / \zeta = 0.31 \%$ and $\delta \Sigma / \Sigma = 0.44 \%$.

As a second example, we obtain errors for a particular phenomenological time-dependent parametrization. 
Following \cite{Simpson} we consider 
$(\mu_\theta(a),\zeta(a), \Sigma(a))$ described by

\begin{equation}\label{s6.2.9}
\mu_\theta (a) = 1 + \left( \mu^0_\theta - 1 \right) \, \frac{1-\Omega_m (a)}{1-\Omega_m},
\end{equation}
\begin{equation}\label{s6.2.11}
\zeta (a) = 1 + \left( \zeta_0 - 1 \right) \, \frac{1-\Omega_m (a)}{1-\Omega_m},
\end{equation}
\begin{equation}\label{s6.2.12}
\Sigma (a) = 1 + \left( \Sigma_0 - 1 \right) \, \frac{1-\Omega_m (a)}{1-\Omega_m},
\end{equation}

\begin{figure} 
  	\includegraphics[width=0.52\textwidth]{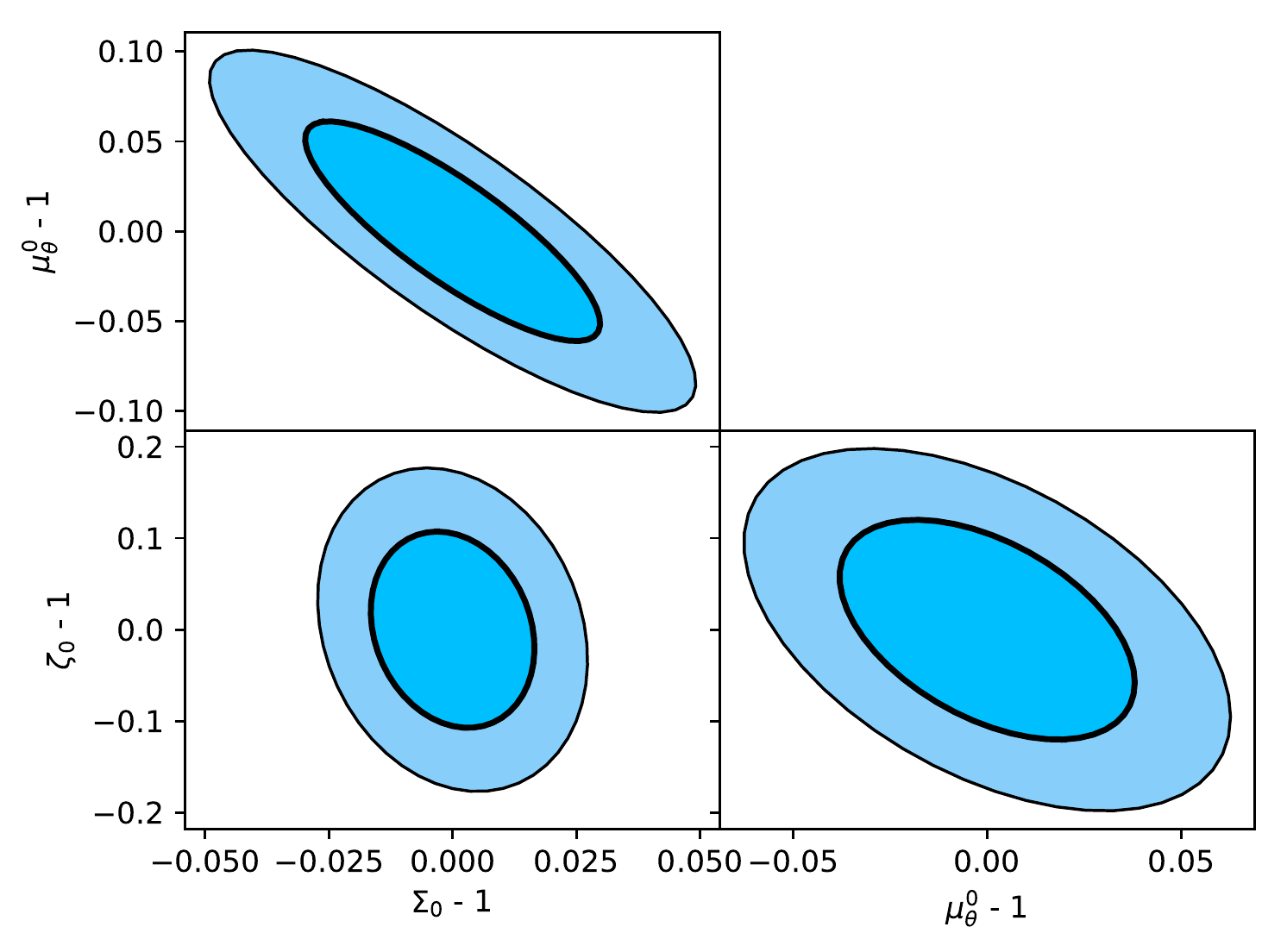}
		\caption{Regions of 1 and 2 sigmas for $\mu_{\theta}^0$, $\zeta_0$ and $\Sigma_0$ of expressions (\ref{s6.2.9},\ref{s6.2.11},\ref{s6.2.12}) using clustering and lensing information in Euclid survey.}
  \label{Figure_2b}
\end{figure}

For a small deviation from $\mathrm{\Lambda CDM}$ we find, using equation (\ref{s4.13}),

\begin{widetext}

\begin{eqnarray}\label{s6.2.10}
f(z) =  \left[ 1 + \left( \zeta_0 - 1 \right) \, \frac{5}{2} \, \left( 1+z \right)^{5/2} \, \int_z^{z_{mat}} \, \frac{1-\Omega_m (z')}{1-\Omega_m}  \, \left( 1+z' \right)^{-7/2} \, dz'   \right] \, \Omega_m (z)^{\hat\gamma},
\end{eqnarray}

\end{widetext}
so we obtain the forecast for $\left( \Sigma_0, \zeta_0, \mu_{\theta}^0 \right)$. We find $\delta \Sigma_0 / \Sigma_0 = 2.21 \%$, $\delta \zeta_0 / \zeta_0 = 8.90 \%$ and $\delta \mu_{\theta}^0 / \mu_{\theta}^0 = 5.08 \%$. We plot the 1 and 2 $\sigma$ regions for $\Sigma_0$, $\zeta_0$ and $\mu_{\theta}^0$ in Fig. \ref{Figure_2b}. 

%

\subsection{The WALLABY survey} \label{sec6.1}
As we have seen, Euclid will not provide accurate measurements of  peculiar velocities. As an example of a future peculiar velocity survey we will consider the SKA-WALLABY survey. 
WALLABY \cite{Johnston:2008hp} (Widefield ASKAP L-band Legacy All-sky Blind surveY) is one of the  ASKAP Survey Science Projects and is focused on enhancing our understanding of the extragalactic neutral hydrogen (HI) universe. It will be able to map the galaxy distribution and galaxy velocity distribution up to redshift $z = 0.26$. This survey will measure $P_{gg}$, $P_{uu}$ and $P_{gu}$ power spectra. The fraction of the sky, the bias and the $k_{\text{max}}$ are respectively $f_{sky} = 0.75$, $b = 0.7$ and $k_{\text{max}} = 0.2$ $h$/Mpc. For $k_{\text{min}}$ we take the value $k_{\text{min}} = 0.007$ $h$/Mpc \cite{Amendola:2013qna}. As independent parameters we consider $(\mu_{\theta}, \zeta, E)$ in each redshift bin. We summarize the redshift bins, galaxy densities and the relative precision for each parameter in Table \ref{Table2}. As we can see, due to the low galaxy densities at low redshift, WALLABY is not competitive measuring modified gravity and imperfect fluid effects. Although peculiar velocity power spectrum could be an interesting observable, current and future surveys will not have enough precision for a competitive measurement of the effective parameters. 

\begin{table}[htbp]
\begin{tabular}{|c|c|c|c|c|}
\hline
$z$ & $n_g \, \times \, 10^{-3}$ & $n_u \, \times \, 10^{-3}$ & $\Delta \mu_{\theta} / \mu_{\theta} (\%)$ & $\Delta \zeta / \zeta (\%)$  \\
\hline \hline 
0.0175  &  67.4  &  8.50    &  2130 &  2130 \\ \hline
0.0350  &  23.1  &  1.06    &  494  &  492 \\ \hline
0.0525  &  8.59  &  0.15    &  870  &  865 \\ \hline
0.0700  &  3.00  &  0.031   &  510  &  506 \\ \hline
0.0875  &  1.09  &  0.0026  &  342  &  340 \\ \hline
0.105   &  0.45  &  0.00097 &  308  &  307 \\ \hline
\end{tabular}
\caption{Redshift bins, galaxy densities in $(\textrm{h/Mpc})^{3}$ and forecasted relative errors for $\mu_{\theta}$ and $\zeta$ for the WALLABY survey.}
\label{Table2}
\end{table}
%

\section{Imperfect dark matter with shear viscosity: present constraints and  forecasts} \label{sec7}

Finally, as an example, we analyse a particular model for dark matter with shear viscosity. We will obtain current  constraints using SDSS data and compare them with the expected precision of Euclid. We consider the particular model of shear viscosity (\ref{s3.3}) where we define the dimensionless parameter $\tilde{\eta} \equiv H_0 \eta_0 / 24 \pi G$ \cite{Barbosa:2017ojt}. We consider the case in which this parameter is constant. In that situation we have found a good analytical approximation for the growth function $f(z)$,

\begin{equation}\label{s7.1}
f(z) = \left( 1 + \frac{a_1}{\left[ 1 + a_2 z + a_3 z^2 \right]^{3/2}} \right) \, \Omega_m^{\hat\gamma} (z),
\end{equation}
\begin{equation}\label{s7.2}
a_1 = \exp \left[ - 0.146 \,\, \hat{\eta}^{0.948} \right] -1,
\end{equation}
\begin{equation}\label{s7.3}
a_2 =  1.447 - 0.106 \, \hat{\eta} + 0.003 \, \hat{\eta}^2,
\end{equation}
\begin{equation}\label{s7.3.1}
a_3 =  0.429 - 0.014 \, \hat{\eta} + 0.0004 \, \hat{\eta}^2,
\end{equation}
being $\hat{\eta} = \tilde{\eta} \, ( k / H_0 )^2$. We will use SDSS matter power spectrum from luminous red galaxies data \cite{Tegmark:2006az} to constrain $\tilde{\eta}$. The observable is the galaxy power spectra today $P^{LRG}(k)$,

\begin{equation}\label{s7.4}
P^{LRG}(k) = b^2 \, \left( \frac{\delta (0)}{\delta (z_{mat})} \, \frac{\delta_{\Lambda} (z_{mat})}{\delta_{\Lambda} (0)} \right)^2 \, P_{\Lambda} (k), 
\end{equation}
where $\delta_{\Lambda} (z)$ and $P_{\Lambda} (k)$ are the growth factor and the matter power spectrum in $\mathrm{\Lambda CDM}$ respectively, and,

\begin{equation}\label{s7.5}
\frac{\delta (0)}{\delta (z_{mat})} \, \frac{\delta_{\Lambda} (z_{mat})}{\delta_{\Lambda} (0)} =  \exp \left[ \int_{N_{mat}}^0 \left( f(N') - f_{\Lambda} (N') \right) \, dN' \right],
\end{equation}
where $N = - \ln (1+z)$ and $N_{mat} \approx -3$. The observational data of $P^{LRG}(k)$ with errors are in Table \ref{Table4}.

\begin{table}[htbp]
\begin{tabular}{|c|c|c|}
\hline
$k$ & $P$ & $\Delta P$ \\
\hline \hline 
 0.012 & 124884 & 18775  \\ \hline
 0.015 & 118814 & 29400  \\ \hline
 0.018 & 134291 & 21638  \\ \hline
 0.021 & 58644  & 16647  \\ \hline
 0.024 & 105253 & 12736  \\ \hline
 0.028 & 77699  & 9666   \\ \hline  
 0.032 & 57870  & 7264   \\ \hline
 0.037 & 56516  & 5466   \\ \hline
 0.043 & 50125  & 3991   \\ \hline
 0.049 & 45076  & 2956   \\ \hline
 0.057 & 39339  & 2214   \\ \hline
 0.065 & 39609  & 1679   \\ \hline
 0.075 & 31566  & 1284   \\ \hline
 0.087 & 24837  & 991    \\ \hline
\end{tabular}
\caption{SDSS luminous red galaxies data \cite{Tegmark:2006az}: $k$ values in $\textrm{h/Mpc}$, LRG power spectrum and errors, both in $(\textrm{Mpc/h})^{3}$.}
\label{Table4}
\end{table}
\begin{figure} 
  	\includegraphics[width=0.485\textwidth]{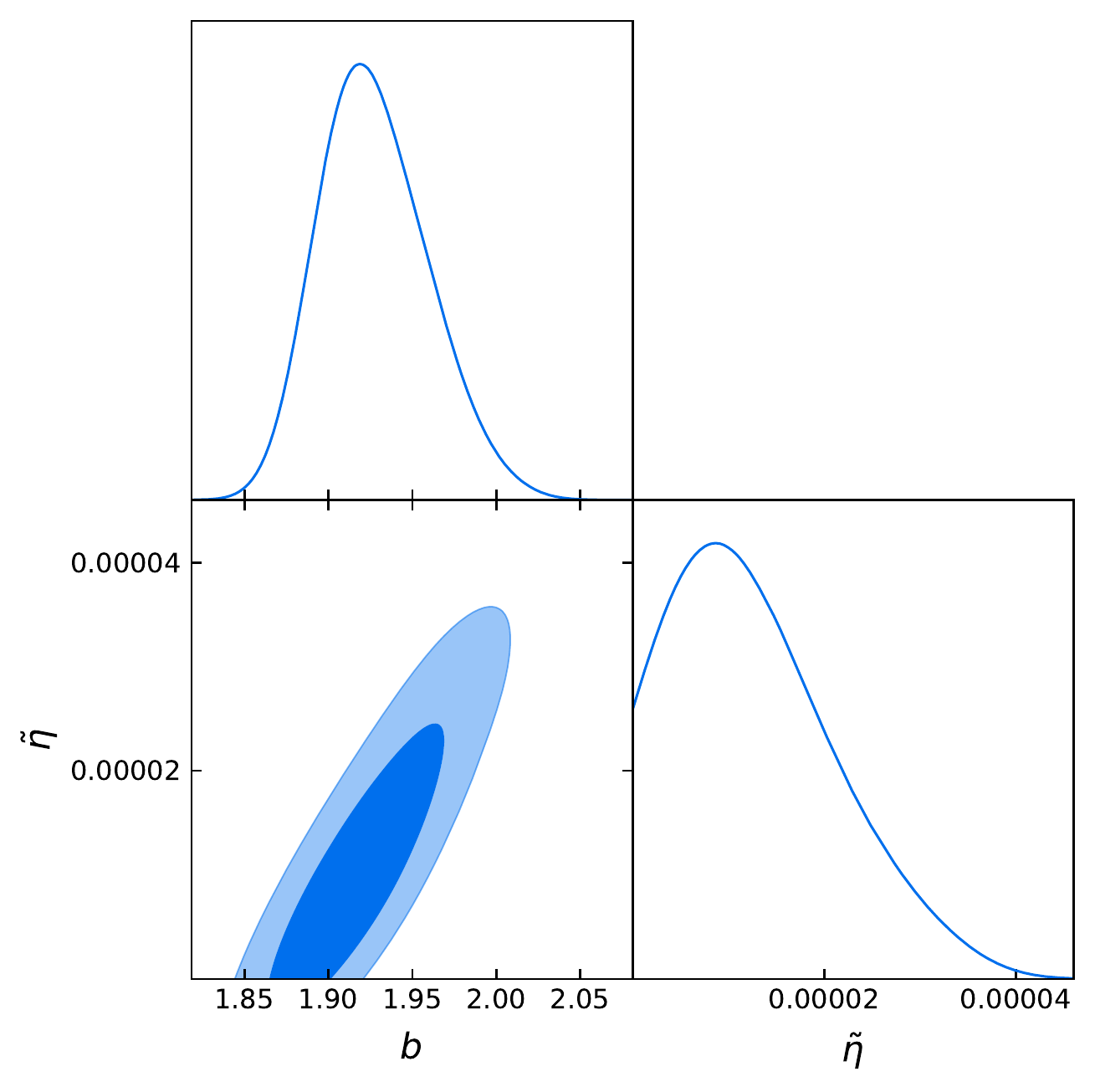}
		\caption{Likelihood and confidence levels for the shear viscosity model using SDSS luminous red galaxies and considering $b$ and $\tilde{\eta}$ as free parameters.}
  \label{Figure_2}
\end{figure}
\begin{figure} 
  	\includegraphics[width=0.495\textwidth]{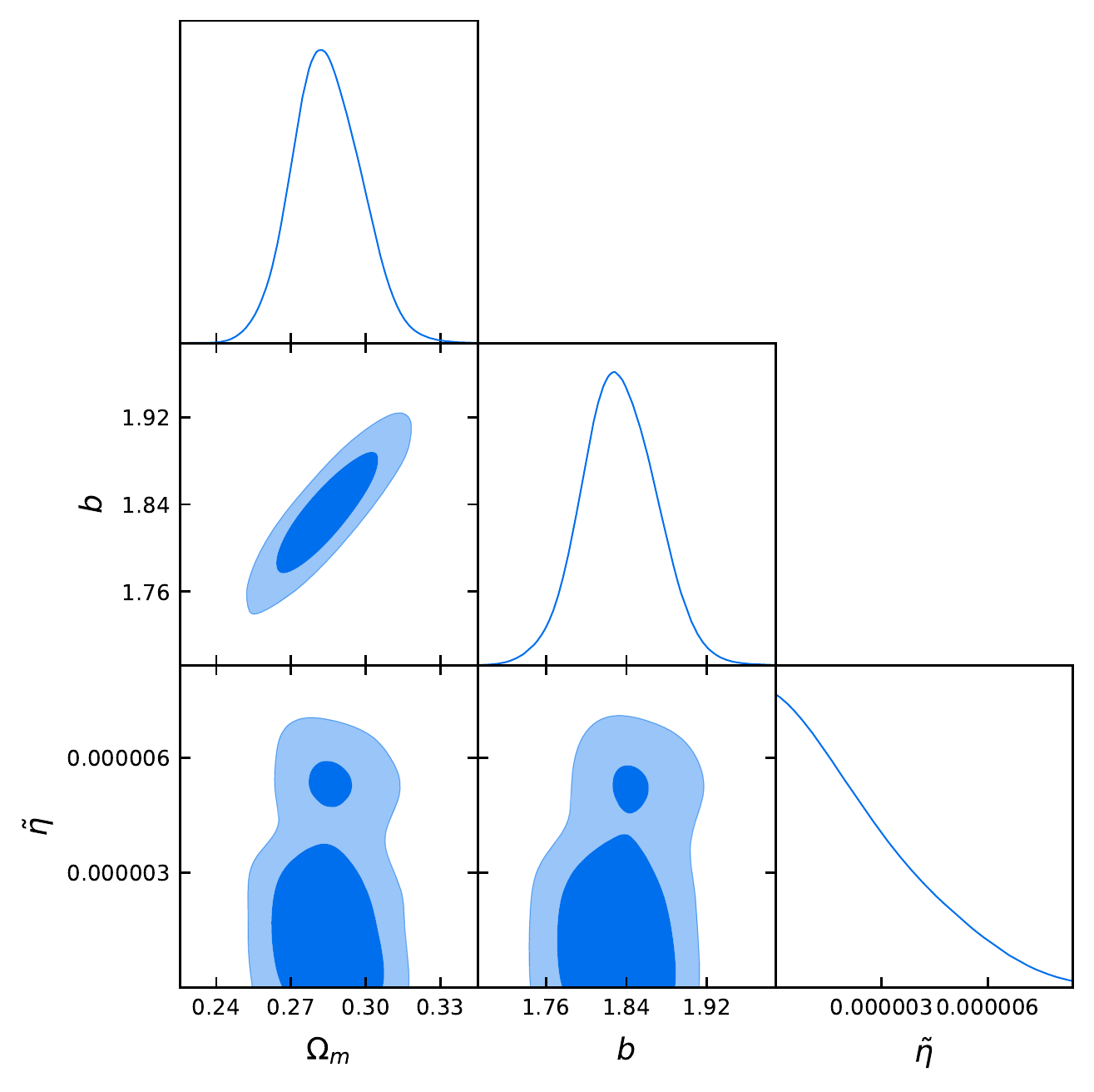}
		\caption{Likelihood and confidence levels for shear viscosity model using SDSS luminous red galaxies and considering $\Omega_m$, $b$ and $\tilde{\eta}$ as free parameters.}
  \label{Figure_3}
\end{figure}

We compute the corresponding $\chi^2$ for expression (\ref{s7.4}) and obtain the best fit and the confidence regions. First of all we consider parameters $(b, \tilde{\eta})$ in Fig. \ref{Figure_2}, fixing the rest of parameters to the fiducial ones. The best fit corresponds to the values $b = 1.916_{-0.051}^{+0.055}$ and $\tilde{\eta} = \left( 0.953^{+1.579}_{-0.953} \right) \times 10^{-5}$. We see that although the best fit corresponds to a non-vanishing viscosity, it is compatible with zero within one sigma. As a matter of fact, we find  $\tilde{\eta}< 3.71 \times 10^{-5}$ at $95 \%$ C.L. 

We have also considered $(\Omega_m, b, \tilde{\eta})$ as fitting parameters in Fig. \ref{Figure_3}. In this case, the best fit corresponds to $\Omega_m = 0.280^{+0.027}_{-0.019}$, $b = 1.82^{+0.08}_{-0.05}$, $\tilde{\eta} = \left( 0.261^{+6.875}_{-0.261} \right) \times 10^{-6}$, which again is compatible with vanishing viscosity at the one-sigma level. We find $\tilde{\eta}<7.55 \times 10^{-6}$ at $95 \%$ C.L.

Finally, we forecast the precision for the future measurements of $\tilde{\eta}$ with Euclid. We compute the clustering and lensing power spectra and obtain the Fisher matrices (\ref{s6.5}) and (\ref{s6.8}) using the information of \ref{sec6.2} and considering $\mathrm{\Lambda CDM}$ as the fiducial model. We summarize the results in Table \ref{Table5}.

\begin{table}[htbp]
\resizebox{9cm}{!} {
\begin{tabular}{|c|c|c|c|c|c|c|}
\hline
$z$ & $k_{\text{max}}$ & $\ell_{\text{max}}$ & $n \, \times \, 10^{-3}$ & $\Delta \tilde{\eta}^C \, \times \, 10^{-7}$ & $\Delta \tilde{\eta}^L \, \times \, 10^{-7}$ & $\Delta \tilde{\eta}^T \, \times \, 10^{-7}$  \\ 
\hline \hline 
0.60  & 0.195 & 300  &  3.56  &  2.53  &  8.18  &  2.41 \\ \hline 
0.80  & 0.225 & 437  &  2.42  &  1.55  &  7.00  &  1.51 \\ \hline  
1.00  & 0.260 & 597  &  1.81  &  1.15  &  8.14  &  1.14 \\ \hline 
1.20  & 0.299 & 782  &  1.44  &  1.05  &  13.1  &  1.04 \\ \hline
1.40  & 0.343 & 994  &  0.99  &  0.97  &  30.6  &  0.97 \\ \hline
1.80  & 0.447 & 1510 &  0.33  &  0.94  &  257   &  0.94 \\ \hline
\end{tabular}
}
\caption{Redshift bins, $k_{\text{max}}$ in $\textrm{h/Mpc}$, $\ell_{\text{max}}$ values, galaxy densities in $(\textrm{h/Mpc})^{3}$ and errors for $\tilde{\eta}$ for the Euclid forecast. Super-index $C$ denotes clustering information, $L$ denotes lensing information and $T$ denotes clustering + lensing information.}
\label{Table5}
\end{table}

As we can see, Euclid improves $1-2$ orders of magnitude the accuracy of $\tilde{\eta}$ with respect to SDSS luminous red galaxies.

\section{Analysis of the results and conclusions} \label{sec8}

In this work we propose a model-independent parametrization of modified gravity and imperfect dark matter perturbations evolution in the QSA and sub-Hubble approximation. Unlike the perfect fluid case \cite{Silvestri:2013ne} in which two parameters $(\mu,\gamma)$ are needed to modify gravity equations (\ref{s1.12} - \ref{s1.13}), three additional parameters are included to  modify the conservation equations for dark matter perturbations (\ref{s2.1} - \ref{s2.2}). Then the complete system of equations is described with five independent parameters $(\mu, \gamma, \mu_m, \mu_d, \mu_{\theta})$ defined in the following way,

\begin{eqnarray}\label{s8.1}
k^2 \, \Phi = -\frac{3}{2} \, {\cal H}^2 \, \Omega_m (a) \mu \, \gamma \, \delta,
\end{eqnarray}
\begin{eqnarray}\label{s8.2}
k^2 \, \Psi = -\frac{3}{2} \, {\cal H}^2 \, \Omega_m (a) \mu \, \delta,
\end{eqnarray}
\begin{eqnarray}\label{s8.3}
\delta'' + {\cal H} \, \mu_d \, \delta' - \frac{3}{2} \, {\cal H}^2 \, \Omega_m (a) \, \mu_m \, \delta = 0,
\end{eqnarray}
\begin{eqnarray}\label{s8.4}
\theta = - \mu_{\theta} \, \delta'.
\end{eqnarray}

This parameterization reduces to $\mathrm{\Lambda CDM}$ when $\mu = \gamma = \mu_m = \mu_d = \mu_{\theta} = 1$. We have proved that a general non-conserved and imperfect fluid for dark matter can be described with this parameterization. Then we study the galaxy survey observables: power spectrum of galaxy distribution $P_{g g}$, peculiar velocities $P_{u u}$, the cross-relation $P_{g u}$; and finally the convergence power spectrum of weak lensing $P_{\kappa\kappa}$. Considering that $\mu_m$ and $\mu_d$ are approximately constant we find a simple parameterization of the growth function $f(z)$; and using this growth function we obtain simple expressions for the power spectra observables as a function of $(\mu, \gamma, \mu_m, \mu_d, \mu_{\theta})$. Then we find that the observables only depend of a reduced subset of parameters $(\Sigma, \zeta, \mu_{\theta})$ which are defined as, 

\begin{equation}\label{s8.5}
\Sigma = \frac{\mu \, (1+\gamma)}{2},
\end{equation}
\begin{equation}\label{s8.6}
\zeta = \frac{1}{4} \left[ \sqrt{(1 - 2 \mu_d)^2+24\mu_m} + 1 - 2 \mu_d \right],
\end{equation}
together with $\mu_{\theta}$ defined in (\ref{s8.4}). This is interesting because a measurement of these parameters can give us some clues about the underlying theory. We explore all the possibilities in Table \ref{Table1}.  There are two cases in which we would extract a lot of information. If $\zeta = \mu_{\theta} = 1$ but $\Sigma \neq 1$ i.e. we measure standard galaxy and peculiar velocity power spectra but a non-standard convergence power spectrum, this can only  be generated by a modified gravity with $\mu = 1$ but $\gamma \neq 1$. If $\Sigma = \mu_{\theta} = 1$ but $\zeta \neq 1$ i.e. we measure standard power spectra but with a non-standard growth function, this can be only generated by a modified dark matter theory with bulk and shear viscosity. More complicated situations produce a degeneration between underlying theories as we can see in Table \ref{Table1}.

 We have also shown that an Euclid-like survey could measure $(\Sigma, \zeta, \mu_{\theta})$ with accuracy of order $1 \, \%$ as we can see in Table \ref{Table3} and Fig. \ref{Figure_2a}. However, as shown in section \ref{sec6} peculiar velocity surveys will not be competitive measuring these parameters.

Finally, in section \ref{sec7} we consider a particular model of modified dark matter with a shear viscosity component \cite{Barbosa:2017ojt}. We use SDSS luminous red galaxies data \cite{Tegmark:2006az} to constrain this model and we obtain that $\bar{\eta} < 7.55 \times 10^{-6}$ at the 95\% C.L. Then we perform the Fisher analysis for the Euclid survey and we find errors for $\bar{\eta}$ of order $10^{-7}$, so new galaxy surveys will improve the accuracy in $1-2$ orders of magnitude.

In conclusion, the phenomenological description introduced in this work can be a useful tool to describe a broad class of non-standard dark matter models. The new effective parameters are particularly suited for future galaxy surveys observations and could help discriminate different dark sector models.

\vspace{0.2cm}
{\it Acknowledgements}. M.A.R acknowledges support from UCM predoctoral grant. This work has been supported by the Spanish  MINECO grants FIS2016-78859-P(AEI/FEDER, UE) and Red Consolider MultiDark FPA2017-90566-REDC.


\end{document}